\documentclass[aps,physrev,preprint,groupedaddress]{revtex4-2}

\usepackage{graphicx}
\usepackage{dcolumn}

\usepackage{siunitx}
\usepackage{braket} 
\usepackage{amsmath}
\usepackage{amssymb}
\usepackage{bbold}
\usepackage{bm}
\usepackage{xcolor}

\usepackage{hyperref}

\usepackage{subcaption}

\newcommand{\nonsymerr}[3]{#1^{#2}_{#3}}

\begin{document}

\preprint{KEK-CP-0411}
\preprint{CERN-TH-2026-093}

\title{Inclusive semileptonic $D_s\to X_s\ell\bar\nu$ decays from lattice QCD: continuum and chiral extrapolation}


\author{Ryan Kellermann}
\email[]{kelry@post.kek.jp}
\affiliation{High Energy Accelerator Research Organization (KEK), Ibaraki 305-0801, Japan}
\author{Alessandro Barone}
\affiliation{PRISMA++ Cluster of Excellence \& Institut f\"ur Kernphysik, Johannes-Gutenberg-Universit\"at Mainz, D-55099 Mainz, Germany}
\author{Ahmed Elgaziari}
\affiliation{School of Physics and Astronomy, University of Southampton, Southampton SO17 1BJ, UK}
\affiliation{STAG Research Center, University of Southampton, Southampton SO17 1BJ, UK}
\author{Shoji Hashimoto}
\affiliation{School of High Energy Accelerator Science, SOKENDAI (The Graduate University for Advanced Studies), Ibaraki 305-0801, Japan}
\affiliation{High Energy Accelerator Research Organization (KEK), Ibaraki 305-0801, Japan}
\author{Zhi Hu}
\affiliation{RIKEN Center for Computational Science (R-CCS), Kobe 650-0047, Japan}
\affiliation{High Energy Accelerator Research Organization (KEK), Ibaraki 305-0801, Japan}
\author{Andreas J\"uttner}
\affiliation{CERN, Theoretical Physics Department, Geneva, Switzerland}
\affiliation{School of Physics and Astronomy, University of Southampton, Southampton SO17 1BJ, UK}
\affiliation{STAG Research Center, University of Southampton, Southampton SO17 1BJ, UK}
\author{Takashi Kaneko}
\affiliation{School of High Energy Accelerator Science, SOKENDAI (The Graduate University for Advanced Studies), Ibaraki 305-0801, Japan}
\affiliation{High Energy Accelerator Research Organization (KEK), Ibaraki 305-0801, Japan}

\date{\today}

\begin{abstract}

We present results for the inclusive semileptonic $D_s \to X_s \ell\bar\nu$ decay rate from lattice QCD. 
Chiral and continuum extrapolations are performed using  gauge ensembles generated with 2+1 flavours of M\"obius domain-wall fermions. Systematic errors are fully addressed including those from the integral over all possible final states. Our results are in agreement with currently available experimental data, with an error at the few--percent level.
\end{abstract}


\maketitle

\section{\label{sec:Introduction}Introduction}

The long-standing tension between determinations of the Cabibbo-Kobayashi-Maskawa (CKM) matrix elements $|V_{cb}|$ and $|V_{ub}|$ from the inclusive and exclusive measurements of $B_{(s)}$ meson decays is limiting the precision of important Standard Model (SM) predictions. For instance, $|V_{cb}|$ plays a crucial role in the SM prediction for neutral-kaon mixing, which in turn is a formidable probe for NP at very high energy scales. But, the tension on $|V_{cb}|$ is the dominant uncertainty in its error budget~\cite{Brod:2019rzc,FlavourLatticeAveragingGroupFLAG:2024oxs}. To resolve the tension, and thus improve various SM tests, we need to strengthen our understanding of the theoretical and experimental uncertainties, both quantitatively and qualitatively.

In this context, while lattice-QCD predictions for the exclusive decay rate have been possible for quite some time (see~\cite{FlavourLatticeAveragingGroupFLAG:2024oxs}), there has recently been progress towards predicting also inclusive decay rates within the same theoretical approach. Although determining and summing all exclusive decay modes to determine the inclusive decay rate remains out of reach in lattice QCD, a technique to directly compute the inclusive decay rate has been developed, where the sum over a given energy range is reconstructed from the Euclidean correlation functions at various time separations. The first results for the inclusive semileptonic process $D_s\to X_s\ell\bar\nu$ have recently been presented, based on Chebyshev reconstruction~\cite{Barata:1990rn,Gambino:2020crt,Bailas:2020qmv} for a single lattice spacing and sea-quark mass~\cite{Kellermann:2025pzt}. An alternative approach, Hansen-Lupo-Tantalo (HLT) reconstruction \cite{Hansen:2019idp,Gambino:2022dvu, DeSantis:2025lal}, was used in \cite{DeSantis:2025qbb, DeSantis:2025yfm} to present a determination of the inclusive rate of the $D_s$ meson including the Cabibbo--suppressed channels proportional to $|V_{us}|$ and $|V_{cd}|$ including the chiral and continuum extrapolations. For a comparison between the Chebyshev approximation and the HLT reconstruction, and work on the inclusive $B_s\to X_c\ell\bar\nu$ decay, we refer to \cite{Barone:2023tbl}.

Our previous work~\cite{Kellermann:2025pzt} concentrated on systematic effects related to the Chebyshev reconstruction, in particular, the bias due to the truncation of the Chebyshev expansion, as well as the finite-volume effect. Here, we investigate the chiral and continuum limits for calculating the inclusive decay rate. The results obtained in this work are complementary to \cite{DeSantis:2025qbb}, using an independent method of the reconstruction and the error analysis. We supplement our previous study with additional gauge ensembles at different pion masses and lattice spacings, in order to extrapolate towards the physical pion mass, as well as the continuum limit.

Our simulations employ M\"obius domain-wall fermions \cite{Shamir:1993zy,Furman:1994ky,Brower:2012vk} for all quarks, including the $c$ and $s$ quarks near their physical values, on lattice ensembles generated for the study of $B$ meson semileptonic decays \cite{Colquhoun:2022atw,Aoki:2023qpa}.

The remainder of this paper is structured as follows. In Section~\ref{sec:Review} we briefly review the necessary theoretical framework introduced in \cite{Kellermann:2025pzt}, recalling the definition of key quantities required for our calculations. Section~\ref{sec:GaugeEnsembles} presents details of the simulation. In Section~\ref{sec:FittingStrategy} we then define the global fitting strategy employed to parametrize the pion mass, lattice spacing and $\bm{q}^2$ dependencies. The application of this fit on the dominant contribution to the inclusive rate is also discussed in this section. Finally, we then use the fitted parameters to determine the inclusive rate in Section~\ref{sec:InclusiveRate}, before presenting a summary and future prospects in Section~\ref{sec:Conclusion}.

\section{\label{sec:Review}Review of inclusive decays}

We provide a brief overview of the key concepts required for the computation of the inclusive decay rate in lattice QCD. For more details we refer the reader to earlier work~\cite{Barone:2023tbl,Kellermann:2025pzt}.

We start with the differential decay rate, which, for the process $D_s \rightarrow X_s \ell\nu_\ell$ studied in this work, is given by 
\begin{align}
  \frac{d\Gamma}{dq^2 dq_0 dE_\ell} = \frac{G_F^2 |V_{cs}|^2}{8\pi^3} L_{\mu\nu} W^{\mu\nu} \, ,
  \label{equ:DiffRateInclusive}
\end{align}
and we neglect QED corrections. The kinematical factors are collected in the leptonic tensor $L_{\mu\nu}$ given by
\begin{align}
  L^{\mu\nu} = p_\ell^\mu p_{\nu_\ell}^\nu + p_\ell^\nu p_{\nu_\ell}^\mu - g^{\mu\nu} p_\ell \cdot p_{\nu_\ell} - i\epsilon^{\mu\alpha\nu\beta} p_{\ell, \alpha} p_{\nu_\ell, \beta} \, .
  \label{equ:LeptonTensor}
\end{align}
Here, $p_\ell \text{ and } p_{\nu_\ell}$ denote the four-momenta of the lepton and neutrino, respectively and the momentum transfer between initial and final states is $q = p_\ell + p_{\nu_\ell}$.

Working in the rest frame of the initial $D_s$ meson, the hadronic contributions are encoded in the hadronic tensor defined as
\begin{align}
  W_{\mu\nu}(\bm{q},\omega) = \frac{1}{2m_{D_s}} \sum_{X_s} \delta(\omega - E_{X_s}) \braket{D_s|\tilde{J}_\mu^\dagger(\bm{q},0)|X_s} \braket{X_s|\tilde{J}_\nu(\bm{q},0)|D_s} \, ,
  \label{equ:HadronicTensor3}
\end{align}
which depends on the kinematic variables $\bm{q}^2 = q_0^2 - q^2$ and $\omega = m_{D_s} - q_0$, {\it i.e.} the three-momentum and energy of the final hadronic state $X_s$. The charged $V-A$ currents (transform to the momentum space) $\tilde J_\mu$ mediate the $c\to s$ flavour-changing transition. As discussed in detail in~\cite{Kellermann:2025pzt}, the hadronic tensor enters the definition of the total decay rate, which we write as
\begin{align}
  \Gamma = \frac{G_F^2 |V_{cs}|^2}{24\pi^3} \int_0^{\bm{q}_{\text{max}}^2} d\bm{q}^2 \sqrt{\bm{q}^2} \bar{X}(\bm{q}^2) \, ,
  \label{equ:TotalRateInclusive}
\end{align}
where the upper limit of the integral is given by $\bm{q}_{\text{max}}^2 = (m_{D_s}^2 - m_{\eta_s}^2)^2/(4M_{D_s}^2)$. The function $\bar{X}(\bm{q}^2)$ contains an integral over the final-state energy $\omega$,
\begin{align}
  \bar{X}(\bm{q}^2) = \sum_{l=0}^{2} \bar{X}^{(l)}(\bm{q}^2) \, , \quad \bar{X}^{(l)}(\bm{q}^2) \equiv \int_{\omega_{\text{min}}}^{\omega_{\text{max}}} d\omega \, X^{(l)}(\bm{q}^2, \omega) \, ,
  \label{equ:XDecomposition}
\end{align}
with integration limits $\omega_{\text{min}} = \sqrt{m_{\eta_s}^2 + \bm{q}^2}$ and $\omega_{\text{max}} = m_{D_s} - \sqrt{\bm{q}^2}$,
where the relation to the hadronic tensor $W^{\mu\nu}$ is given as
\begin{align}
  \begin{split}
    X^{(0)}(\bm{q}^2, \omega) &= \bm{q}^2 W_{00} + \sum_{i} (q_i^2 - \bm{q}^2) W_{ii} + \sum_{i \neq j} q^{i} W_{ij} q^{j} \, , \\
    X^{(1)}(\bm{q}^2, \omega) &= -q_0 \sum_{i} q^{i} \left(W_{0i} + W_{i0}\right) \, ,\\
    X^{(2)}(\bm{q}^2, \omega) &= q_0^2 \sum_{i} W_{ii} \, .\\
  \end{split}
  \label{equ:XComponents}
\end{align}
As these formulae show, the total decay rate is obtained from an integral over the energy $\omega$ of the hadronic final state, as well as the corresponding three momentum $\bm{q}^2$. In essence, $X^{(l)}(\bm{q}^2, \omega)$ encodes all information on the hadronic dynamics as a linear combination of the hadronic tensors with some kinematical factors. 

Given the $V-A$ nature of the charged current in $X^{(l)}$, the hadronic tensor and the corresponding $X^{(l)}(\bm{q}^2,\omega)$ can be written as
\begin{align}
  X^{(l)} = X^{(l), VV} + X^{(l), AA} - X^{(l), VA} - X^{(l), AV} \, ,
  \label{equ:Decomposition}
\end{align}
where $V$ and $A$ denote the insertion of the vector and axial-vector currents, respectively. The decomposition for $\bar{X}^{(l)}$ follows in a similar manner. In the massless-lepton limit the contributions from $VA$ and $AV$ to the decay rate vanish due to parity. 

Let us also briefly review the decomposition of $X(\bm{q}^2, \omega)$ into longitudinal and transverse components for the vector ($VV$) or axial-vector ($AA$) insertions of the hadronic tensor, allowing us to identify contributions of different final states. The numerical results discussed in Section \ref{sec:FittingStrategy} will be based on this decomposition. 
We decompose $X(\bm{q}^2,\omega)$, given in Eqs.~\eqref{equ:XDecomposition}--\eqref{equ:XComponents}, as $X(\bm{q}^2,\omega) = X_\parallel(\bm{q}^2,\omega)+X_\perp(\bm{q}^2,\omega)$,  defining the longitudinal and transverse contributions, $X_\parallel(\bm{q}^2, \omega)$ and $X_\perp(\bm{q}^2, \omega)$, respectively, as
\begin{align}
  X_\parallel(\bm{q}^2, \omega) &= \bm{q}^2 W_{00} - q_0 \sum_i q_i (W_{i0} + W_{0i}) + \frac{q_0^2}{\bm{q}^2} \sum_{i,j} q_i W_{ij} q_j \label{equ:ParallelDecomposition} \, , \\
  \begin{split}
    X_\perp(\bm{q}^2, \omega) &= (q_0^2 - \bm{q}^2) \sum_{i,j} \left[ \delta^{ij} - \frac{q_iq_j}{\bm{q}^2}\right] W_{ij}  \\
    &= \left(1 - \frac{q_0^2}{\bm{q}^2}\right) \left[\sum_{i} (q_i^2 - \bm{q}^2) W_{ii} + \sum_{i \neq j} q_i W_{ij} q_j \right] \label{equ:PerpDecomposition} \, .
  \end{split}
\end{align}
In \cite{Kellermann:2025pzt} we showed how these expression can be combined with form factors defined in a way motivated by heavy-quark effective theory (HQET) to estimate the contribution of the lowest-lying energy state to the inclusive decay rate.

We now discuss how the integral in \eqref{equ:XDecomposition} can be obtained from a lattice computation of the 
time dependence of the Euclidean four-point function
\begin{align}
  C^{SJJS}_{\mu\nu}(\bm{q}, t_{\text{snk}}, t_2, t_1, t_{\text{src}}) \stackrel{t_2 \geq t_1}{=} \sum_{\bm{x}_{\text{snk}}, \bm{x}_{\text{src}}} \braket{\mathcal{O}^{S}_{D_s}(x_{\text{snk}}) \tilde{J}_\mu^\dagger(\bm{q},t_2) \tilde{J}_\nu(\bm{q},t_1) \mathcal{O}^{S\dagger}_{D_s}(x_{\text{src}})} \, ,
  \label{equ:FourPointDefinition}
\end{align} 
where $\mathcal{O}_{D_s}^S$ defines an interpolating operator of the quantum numbers of the $D_s$ meson, and $\tilde{J}_\nu(\bm{q}, t) \equiv \sum_{\bm{x}} \exp(-i\bm{q}\cdot\bm{x}) J_\nu(\bm{x},t)$ with $J_\nu=\bar c \gamma_\nu(1-\gamma_5)s$ projects the currents onto a specific three-momentum. Through this setup, a $D_s$ meson that carries zero momentum is created at the source time slice $t_{\text{src}}$, and is annihilated at the sink $t_{\text{snk}}$.

By taking suitable ratios of the four-point function \eqref{equ:FourPointDefinition} with zero--momentum two--point functions and ensuring ${t_{\rm snk}\gg t_2 \geq t_1\gg t_{\rm src}}$ we can construct matrix elements as 
\begin{align}
  C_{\mu\nu}(\bm{q},t) = \frac{1}{2m_{D_s}} \braket{D_s|\tilde{J}_\mu^\dagger(\bm{q},0) e^{-\hat{H}t} \tilde{J}_\nu(\bm{q},0)|D_s} \, ,
  \label{equ:LatticeCorrelatorRatio}
\end{align}
with $t=t_2-t_1$. We refer to \cite{Kellermann:2025pzt} for a more detailed derivation.
The relation to the  hadronic tensor \eqref{equ:HadronicTensor3} is given in terms of the Laplace transform
\begin{align}
  \begin{split}
    C_{\mu\nu}(\bm{q},t) &= \int_0^\infty d\omega\,\frac{1}{2m_{D_s}} \braket{D_s|\tilde{J}_\mu^\dagger(\bm{q},0) \delta(\hat{H} -\omega) \tilde{J}_\nu(\bm{q},0)|D_s} e^{-\omega t} \\
    &= \int_0^\infty d\omega\, W_{\mu\nu} (\bm{q}, \omega) e^{-\omega t} \, .
  \end{split}
  \label{equ:SpectralRepresentation}
\end{align}

To relate the energy integral to a quantity that can be calculated on the lattice we consider a generic way in which $\bar{X}^{(l)}(\bm{q}^2)$ can be written 
\begin{align}
  \begin{split}
    \bar{X}^{(l)}(\bm{q}^2) = \int_{\omega_{0}}^{\infty} d\omega\, W^{\mu\nu}(\bm{q}, \omega) K_{\mu\nu}^{(l)}(\bm{q}, \omega) \, ,
  \end{split}
  \label{equ:EnergyIntegralXBar}
\end{align}
where we defined the kernel function $K_{\mu\nu}^{(l)}(\bm{q},\omega)\equiv k_{\mu\nu}^{(l)}(\bm{q},\omega) \theta(\omega_{\text{max}} - \omega)$, as a product of a known function $k_{\mu\nu}^{(l)}(\bm{q},\omega)$ and a step function $\theta(\omega_{\text{max}} - \omega)$, removing all contributions above $\omega_{\text{max}}$. The lower limit of the integral has been shifted to $\omega_0$, with $\omega_0 \leq \omega_{\rm min}$, as there are no states below $\omega_{\rm min}$ described by the hadronic tensor. The advantages of freely choosing $\omega_0$ have been discussed in \cite{Kellermann:2025pzt, Barone:2023tbl}, and we choose $\omega_0=0.9\omega_{\rm min}$ throughout this work. 


From there, expanding the kernel in polynomials \cite{Gambino:2020crt} of $e^{-a\omega}$ (for simplicity, we set $a = 1$ in the following) up to an order $N$, 
\begin{align}
  K_{\sigma,\mu\nu}^{(l)}(\bm{q}, \omega) \simeq c_{\mu\nu,0}^{(l)} (\bm{q};\sigma) + c_{\mu\nu,1}^{(l)} (\bm{q};\sigma) e^{-\omega} + \cdots + c_{\mu\nu,N}^{(l)} (\bm{q};\sigma) e^{-N\omega} \, ,
\end{align}
our target quantity can be expressed as
\begin{align}
  \begin{split}
    &\bar{X}_{\sigma}^{(l)}(\bm{q}^2) \simeq \int_{\omega_0}^{\infty} d\omega\, W^{\mu\nu}(\bm{q},\omega) e^{-2\omega t_0} K_{\sigma, \mu\nu}^{(l)}(\bm{q}, \omega; t_0) \\
    &\simeq c_{\mu\nu,0}^{(l)}(\bm{q};\sigma; t_0) \int_{\omega_0}^{\infty} d\omega\, W^{\mu\nu}(\bm{q},\omega) e^{-2\omega t_0} + c_{\mu\nu,1}^{(l)}(\bm{q};\sigma; t_0) \int_{\omega_0}^{\infty} d\omega\, W^{\mu\nu}(\bm{q},\omega) e^{-2\omega t_0} e^{-\omega} \\
    &+ \cdots + c_{\mu\nu,N}^{(l)}(\bm{q};\sigma; t_0) \int_{\omega_0}^{\infty} d\omega\, W^{\mu\nu}(\bm{q},\omega) e^{-2\omega t_0} e^{-N\omega} \, .
  \end{split}
  \label{equ:PolynomialApproximationStart}
\end{align}
The subscript $\sigma$ on $\bar{X}$ refers to a smoothening of the step function, which we introduce to better control the polynomial approximation. Namely, we replace the step function by $\theta_{\sigma}(x) = 1/(1 + e^{-x/\sigma})$, where it is understood that the $\sigma\to 0$ limit eventually has to be taken to produce the physical decay rate. Lattice simulations are carried out in finite volume, where the hadronic tensor represents a discrete spectrum. The energy integral together with the leptonic tensor in this sense constitute a smearing. Thanks to the smooth function $\theta_\sigma$, the smearing persists also in the kinematical regime, where, in \eqref{equ:EnergyIntegralXBar}, $\omega_0$ approaches $\omega_{\rm max}$.


To avoid the contact term of $t_1 = t_2$ appearing in \eqref{equ:FourPointDefinition}, which receives contributions from the opposite time ordering corresponding to unphysical $\bar{c}ss\bar{c}$ final states, we introduce the factor $e^{-2\omega t_0}$ which we compensate for in the kernel function $K_{\sigma, \mu\nu}^{(l)}(\bm{q},\omega; t_0) \equiv e^{2\omega t_0} K_{\sigma, \mu\nu}^{(l)}(\bm{q},\omega)$. We choose $t_0 = 1/2$ throughout this paper.
By comparing \eqref{equ:SpectralRepresentation} and \eqref{equ:PolynomialApproximationStart} we obtain
\begin{align}
  \bar{X}_{\sigma}^{(l)}(\bm{q}^2) \simeq \sum_{k=0}^{N} c_{\mu\nu,k}^{(l)}(\bm{q};\sigma; t_0) C^{\mu\nu}(\bm{q}, k+ 2 t_0) \, .
  \label{equ:ApproximationIntegral}
\end{align}
In the continuum limit, the time separation $t_0=a/2$ vanishes. There is a potential divergence when the two currents approach each other. In momentum space, such a divergence is controlled by the momentum $q_\mu$ that flows through the strange quark propagator between the two currents. In our setup, the spatial momentum $\bm{q}$ is fixed during the calculation and finally integrated up to $\bm{q}^2_{\text{max}}$. The temporal component $q_0$ represents the energy $\omega$ flowing through the strange quark propagator, and again it is integrated up to $\omega_{\text{max}}$. In either case, the upper limit is set by the kinematics and the integral does not run to infinity. Therefore, the suspected divergence in the limit of $t_0 \to 0$ does not exist in this calculation.

In this way, an expression relating $\bar{X}_{\sigma}^{(l)}(\bm{q}^2)$ appearing in the smeared total decay rate to a quantity which we can compute on the lattice, {\it i.e.} $C^{\mu\nu}(\bm{q}, t)$, is obtained. It is an approximation up to a finite order $N$, where $N$ corresponds to the maximal Euclidean time separation between the inserted currents in the four-point function \eqref{equ:FourPointDefinition}. The final task is then to perform the phase-space, or $\bm{q}^2$-, integral in \eqref{equ:TotalRateInclusive}, for some values of $\sigma$, and then to take the limit of $\sigma\to 0$.

\section{\label{sec:GaugeEnsembles}Simulation details}
We employ gauge ensembles generated with $2+1$ flavors of dynamical quarks in lattice QCD from the JLQCD collaboration. Using the notation from \cite{Colquhoun:2022atw}, we assign each ensemble an ID using the prescription ``X-$ud$\#-$s$a'', where X $\in \{\mathrm{C},\mathrm{M}\}$ denotes the lattice spacing, the number following $ud$ denotes the pion mass given in units of $\SI{100}{\mega\electronvolt}$ and the letter ``a'' after $s$ denotes that all ensembles used in this work possess a strange quark above its physical value.
The coarse ``C'' and medium ``M'' lattices possess a lattice spacing of $a \simeq \SI{0.080} {\femto\meter}$ and $\SI{0.055} {\femto\meter}$, corresponding to lattice cut-offs $a^{-1}\simeq \SI{2.453(1)}{\giga\electronvolt}$ and $\SI{3.61(1)}{\giga\electronvolt}$, respectively. The latter has been used in our previous analysis \cite{Kellermann:2025pzt}. These parameters are determined through the Yang-Mills gradient flow \cite{BMW:2012hcm}. To achieve better control over the discretization errors we employ the tree-level improved Symanzik gauge action and stout smearing \cite{Morningstar:2003gk} to the gauge field when coupled to fermions. We use the M\"obius domain-wall action \cite{Brower:2012vk, Tomii:2016xiv} for both heavy and light quarks. For additional information on the practical implementation and formulation of the quark action in five dimensions we refer to \cite{Brower:2012vk, Boyle:2015vda, Colquhoun:2022atw, Aoki:2023qpa}. The choice of light-quark masses used in this work corresponds to pion masses ranging from $m_\pi \simeq \SI{500} {\mega\electronvolt}$ ($ud5$) down to $\SI{230} {\mega\electronvolt}$ ($ud2$). The charm masses are fixed to their physical values from the spin-averaged charmonium mass \cite{Nakayama:2016atf}. The lattice volumes have been chosen such that the spatial extent of the lattice is kept constant at $L\sim \SI{2.6}{\femto\meter}$. The only exception to this is the ensemble ``C-$ud2$-$s$a-L'' which has a larger volume with spatial extent $L\sim \SI{3.8}{\femto\meter}$.
All our ensembles satisfy the condition $M_\pi L > 4$, where $L$ is the spatial extent of the lattice. This condition is usually imposed to ensure a sufficient suppression of finite-volume effects to a regime at the few-percent level for meson masses and form factors. 
Due to the finite fifth dimension $L_5$, M\"obius domain-wall fermions do not have exact chiral symmetry. A measure of the violation of chiral symmetry is the residual quark mass, which for the lattice used in this work is below $\SI{0.2} {\mega\electronvolt}$, much smaller than the physical masses of the up and down quarks.

We use the vector-current renormalization constant $Z_V$ from \cite{Tomii:2016xiv, Nakayama:2016atf}, determined for the analysis of the short-distance current correlator of light quarks. Thanks to the approximate chiral symmetry of domain-wall fermions, $Z_V=Z_A$. For the ensembles considered in this work the numerical values are $0.9553(92)$ and $0.9636(58)$ for $a \simeq \SI{0.080}{\femto\meter}$ and $\SI{0.055}{\femto\meter}$, respectively, where statistical and systematic errors have been added in quadrature.

For our simulations, we average over $N_{\text{Conf}}$ statistically independent gauge configurations, performing the measurement for each configuration with $N_{t_{\text{src}}}$ evenly distributed choices of the time source. 
Our choices for $N_{\text{Conf}}$ and $N_{t_{\text{src}}}$ are given in Table~\ref{tab:GaugeEnsembles}. 
We use $\mathbb{Z}_2$ wall sources to improve the signal \cite{Foster:1998vw, McNeile:2006bz, Boyle:2008rh}. To cover the whole kinematical region, we introduce four different momenta in the four-point function \eqref{equ:FourPointDefinition}. In terms of $\bm{q} = (2\pi/L) \bm{n}$, we take $\bm{n} = (0,0,0)$, $(0,0,1)$, $(0,1,1)$, and $(1,1,1)$. The ensemble ``C-$ud2$-$s$a-L'' requires additional momenta $\bm{n} = (0,0,2)$, $(0,1,2)$, $(1,1,2)$. All correlation functions analyzed in this work have been computed using the Grid \cite{BoyleGrid, Boyle:2015tjk, Yamaguchi:2022feu} and Hadrons \cite{PortelliHadrons} software packages. For most of the fits we have employed python packages lsqfit \cite{LepageLSQ, Lepage:2001ym} and corrfitter \cite{LepageCORR}. The gvar class \cite{LepageGVAR} is used to capture statistical correlations between data points as well as correlations between data points and priors, allowing for a straightforward treatment of Gaussian-distributed random variables.

For the computation of \eqref{equ:LatticeCorrelatorRatio}, the ratio between two- and four-point correlation functions, we take a source-sink separation $T = t_{\text{snk}} - t_{\text{src}}$ depending on the temporal extent of the lattice. We further define time-slices $t_1$ and $t_2$, chosen in such a way that a sufficient ground-state saturation is achieved for the $D_s$. The choices for $T$ and $[t_1, t_2]$ are also summarized in Table~\ref{tab:GaugeEnsembles}.

More details on the implementation and simulation setup can be found in \cite{Colquhoun:2022atw, Kellermann:2025pzt}

\begin{table}[tbp]
  \begin{tabular}{l|c c c c c c c}
    Lattice parameters/ID & $m_{ud}$ & $m_{s}$ & $M_{\pi}$ [MeV] & $T$ & $[t_1,t_2]$ & $N_{\text{Conf}}$ & $N_{t_{\text{src}}}$ \\
  \hline
  $\beta=4.17$, $a^{-1}=\SI{2.453(1)}{\giga\electronvolt}$, $32^3 \times 64 \times 12$ & & & & & & & \\ 
   C-$ud3$-$s$a & 0.0070 & 0.0400 & $\num{309(1)}$ & 28 & $[11, 17]$ & 100 & 2 \\
   C-$ud4$-$s$a & 0.0120 & 0.0400 & $\num{399(1)}$ & 28 & $[11, 17]$ & 100 & 2 \\
   C-$ud5$-$s$a & 0.0190 & 0.0400 & $\num{499(1)}$ & 28 & $[11, 17]$ & 100 & 2 \\
  \hline
  $\beta=4.17$, $a^{-1}=\SI{2.453(1)}{\giga\electronvolt}$, $48^3 \times 96 \times 12$ & & & & & & & \\  
  C-$ud2$-$s$a-L & 0.0035 & 0.0400 & $\num{226(1)}$ & 42 & $[16, 26]$ & 100 & 4 \\
  \hline
  $\beta=4.35$, $a^{-1}=\SI{3.61(1)}{\giga\electronvolt}$, $48^3 \times 96 \times 8$ & & & & & & & \\  
  M-$ud3$-$s$a & 0.0042 & 0.0250 & $\num{300(1)}$ & 42 & $[16, 26]$ & 50 & 8  \\
  \hline
  \end{tabular}
  \caption{Simulation parameters. We use the notation $N_s^3 \times N_t \times N_5$ for the five dimensional lattice. Quark masses are the bare values in lattice units.}
  \label{tab:GaugeEnsembles}
\end{table}

\section{\label{sec:FittingStrategy}Chiral and continuum extrapolation}
In this section we describe our strategy to extrapolate towards the continuum limit and the physical pion mass. We perform a global fit parameterizing the $\bm{q}^2$, $m_{\pi}^2$ and $a^2$ dependencies simultaneously. We use the decomposition of $\bar{X}(\bm{q}^2)$ into different channels distinguished by longitudinal and transverse components of the vector (VV) and axial-vector (AA) current insertions introduced in Section \ref{sec:Review}. 


For the $\bm{q}^2$-dependence, we assume that the overall shape is shared among different ensembles and can be described by a single set of parameters, and following \cite{Kellermann:2025pzt}, we employ a polynomial interpolation.  The dependencies on the pion mass and lattice spacing are both expected to be minor corrections of order $\mathcal{O}(m_\pi^2)$ and $\mathcal{O}(a^2)$, respectively. We also include a term proportional to $(a\bm{q})^2$ in the fit, to take momentum-dependent cut-off effects into account. 

The fit prescription is then given by
\begin{align}
	\bar{X}^{\mathcal{J}}_{\mathcal{P}}(\bm{q}^2, m_\pi^2, a^2) = \left(\sum_{i=0}^{N_p} c_{i} (\bm{q}^2)^i\right) \times \left(1 + \tilde{c}_{m_\pi}\left(\frac{m_\pi}{\Lambda_c}\right)^2\right) \times \left(1 + \tilde{c}_{a} \left(a\Lambda_c\right)^2\right) \times \left(1 + \tilde{c}_{a\bm{q}} (a\bm{q})^2\right)  \, ,
        \label{equ:ExtrapolationFit}
\end{align}
where $\mathcal{J}$ and $\mathcal{P}$ specify the inserted currents, $VV$ or $AA$, and their projection, $\parallel$ or $\perp$, respectively. 
The dimension of the coefficients $c_i$ is given by $(\si{\giga\electronvolt}^{1-i})^2$. A mass scale $\Lambda_c$ is introduced, so that the coefficients $\tilde{c}_{m_\pi}$, $\tilde{c}_{a}$ and $\tilde{c}_{a\bm{q}}$ in the correction terms are dimensionless. We choose $\Lambda_c = \SI{0.7}{\giga\electronvolt}$. 

The polynomial order $N_p$ of the interpolation is chosen depending on the available number of data points in each channel, such that the fit degree of freedom is 1. For $\{\mathcal{J}, \mathcal{P}\} = \{VV, \parallel\}$, the lightest hadronic final state is the $\eta_s$ meson~\footnote{This work ignores disconnected strange-quark-loop diagrams, and we label the $s\bar{s}$ pseudoscalar $\eta_s$ instead of $\eta$ or $\eta'$.} and four-momentum points are within the range of decay kinematics. We hence use $N_p=3$ for the interpolation.
All other channels share the vector $\phi$ meson as the lightest hadronic state, and consequently the highest $\bm{q}^2$ value in our data is above the kinematical threshold. Removing one data point from the interpolation, we parametrize with $N_p=2$. We confirmed that the fit with polynomial order $N_p\pm1$ yields results that are consistent within errors. 

In the tables given in the next sections, the fit results are from the ansatz where all terms in \eqref{equ:ExtrapolationFit} are kept. In order to quantify the systematic uncertainty from the continuum and physical pion mass extrapolations, we have repeated the analysis while adjusting the model \eqref{equ:ExtrapolationFit} as follows: (i) including $\mathcal{O}\left((m_\pi/\Lambda_c)^4\right)$, (ii) excluding $(a\Lambda_c)^2$, (iii) excluding $(a\bm{q})^2$, and (iv) excluding higher order terms such as $\mathcal{O}((\bm{q}^2)^3(m_\pi/\Lambda_c)^2(a\Lambda_c)^2)$. We refrain from completely removing the $a^2$ dependence, {\it i.e.} assuming a flat extrapolation, as the data exhibit some dependence, see Fig.~\ref{fig:LSpacingDependence}, although our analysis is limited to only two data points for the relevant extrapolation. 

All fits are performed with correlations taken into account; we use the gvar class~\cite{LepageGVAR} to capture all correlations between data points.
The fits satisfy $\chi^2/\text{d.o.f.} \leq 1$. And, the  systematic error is defined as the largest spread in central values between \eqref{equ:ExtrapolationFit} and the different ansatz listed above. 

In order to improve precision, following \cite{Kellermann:2025pzt}, we parameterize the relevant matrix elements in terms of form factors in HQET to express the correlators making up $\bar{X}(\bm{q}^2, m_\pi^2, a^2)$ in terms of their lowest-lying $S$- and $P$-wave states. We perform a fit with all correlators as input to extract the $S$-wave contribution, which is then treated as exact under the energy integral, while the Chebyshev approximation is applied for the excite-state contributions.
As the ground state gives a dominant  contribution, especially for large recoil momenta, this enables better control of systematic uncertainties. 

For excited-state contributions, we use properties of Chebyshev polynomials 
to estimate the systematic error associated with the truncation of Chebyshev polynomials. Since the available order $N$ of the Chebyshev approximation is limited by a finite number of time slices $t$ as well as the statistical error of the correlator, we assume that, above a certain polynomial order $N_{\text{Cut}}$, the unknown matrix elements in the Chebyshev expansion obey a uniform distribution within $[-1,+1]$. 
We confirmed that the $N\to\infty$ limit (combined with the $\sigma\to 0$ limit through $\sigma=1/N$) is stable among different ensembles and found that the error is well under control and is dominated by statistics.

Following the discussion in~\cite{Kellermann:2024jqg} we estimate finite-volume effects using a model of the dominant two-body $K\bar{K}$ states. The model prescription is given by
\begin{align}
    C(\bm{q}, t) = A_0(\bm{q}) e^{-E_0(\bm{q})t} + \mathcal{S}(\bm{q}) \sum_{\bm{k}} A_{\bm{k}}(\bm{q}) e^{-E_{\bm{k}}(\bm{q}) t} |F(E_{\bm{k}}(\bm{q}))|^2 \, ,
\label{equ:ModelFitFunction}
\end{align}
where the first term represents the ground state. The remaining terms describe the $D_s\to K\bar{K}\ell\bar\nu$ process. They are written in terms of a known energy 
\begin{align}
E_{\bm{k}}(\bm{q})=\sqrt{\bm{k}^2 + m_K^2} + \sqrt{(\bm{q} + \bm{k})^2 + m_K^2}
\end{align}
and an amplitude:\footnote{The amplitude given here differs from \cite{Kellermann:2025pzt}, since we found a mistake in our previous analysis. We have confirmed that the corrected amplitude does not affect the predicted finite-volume corrections from our previous work.}
\begin{align}
A_{\bm{k}}(\bm{q})=\frac{\pi}{V} \frac{(\bm{q} + 2\bm{k})^2}{4\sqrt{\bm{k}^2 + m_K^2}\sqrt{(\bm{q} + \bm{k})^2+ m_K^2}}.
\end{align}
The overall factor $\mathcal{S}(\bm{q})$ is determined from a fit to the numerical data, and is assumed to be volume independent. The function $F(E)=1/(E^2-m_\phi^2)$ assumes the role of a ``form factor'', motivated by the idea of vector-meson dominance,
enhancing states close to the vector-meson resonance. The overall normalization is irrelevant as it is absorbed by $\mathcal{S}(\bm{q})$ in \eqref{equ:ModelFitFunction}. It predicts finite-volume corrections much smaller than the current statistical precision. We nonetheless correct our data points by adding the finite-volume corrections.

\section{Results for differential decay rates}
The numerical results are given in this section. 
We show the values of $\bar{X}(\bm{q}^2, m_\pi^2, a^2)$ renormalized by $Z_V^2$, and after the corrections of $V\to\infty$ and $\sigma\to 0$ taken into account. 
The discussion concentrates on the analysis of 
the dominant contribution, {\it i.e.} $\bar{X}^{VV}_{\parallel}(\bm{q}^2)$.
It carries over in the same way to the other channels.


We fit the lattice data for $\bar{X}^{VV}_{\parallel}(\bm{q}^2)$ to \eqref{equ:ExtrapolationFit}. The data entering the fit, as well as the resulting parameters, are listed in Tables~\ref{tab:XVVParallelDataForFit} and~\ref{tab:XVVParallel_Multifit}, respectively. The data and fit parameters for all other channels can be found in Appendix~\ref{sec:TablesAndFigures}.

\begin{figure}[tbp]
    \centering
    \includegraphics[width=\textwidth]{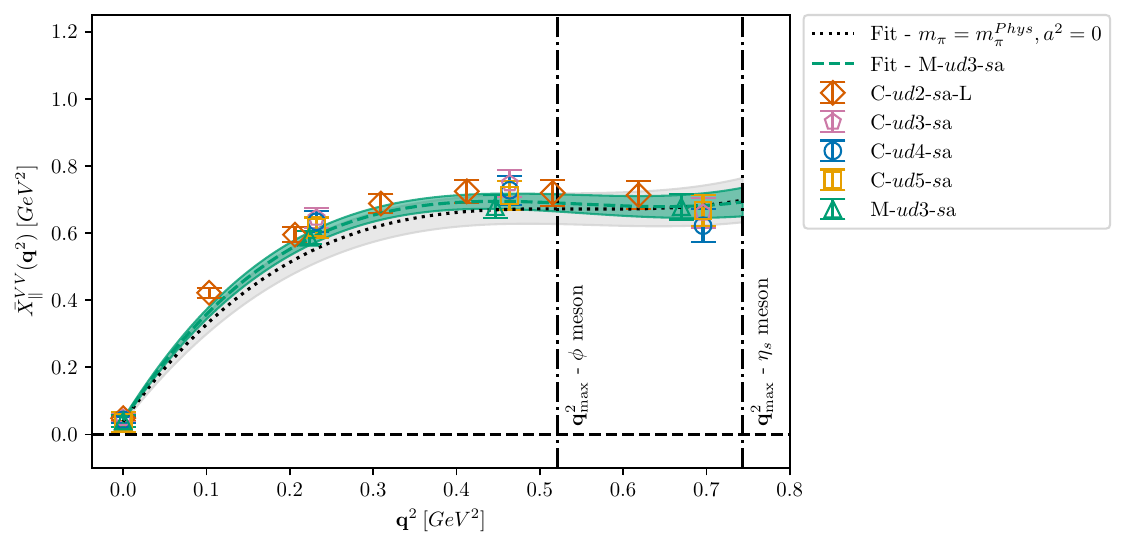}
    \caption{$\bar{X}^{VV}_{\parallel}(\bm{q}^2)$ for all ensembles that enter the fit at all values of $\bm{q}^2$. 
    The dashed line represents the fit result to the ensemble ''M-$ud$3-$s$a'', while the dotted line is extrapolated to the physical-pion-mass and the continuum limit, {\it i.e.} $m_\pi^2 = \left(m_\pi^{\text{Phys}}\right)^2$ and $a^2=0$.}
    \label{fig:XVVParallelFitComparisonPhysiclaPoint}
\end{figure}

In Fig.~\ref{fig:XVVParallelFitComparisonPhysiclaPoint} we plot the data for individual ensembles. The values of $\bm{q}^2$ between different $\beta$ are slightly mismatched, as can be seen in the plot. To verify that the fit describes the data, we include the fit result corresponding to the ensemble M-$ud$3-$s$a 
(triangles) by the dashed line. 
Other data points are also well described by the fit ansatz.
The fit result extrapolated to the continuum and physical pion mass, {\it i.e.} $m_\pi^2 = \left(m_\pi^{\text{Phys}}\right)^2$ and $a^2=0$ is also shown by the dotted line. 

\begin{figure}[tbp]
    \centering
    \begin{subfigure}{0.49\textwidth}
        \centering
        \includegraphics[width=\textwidth]{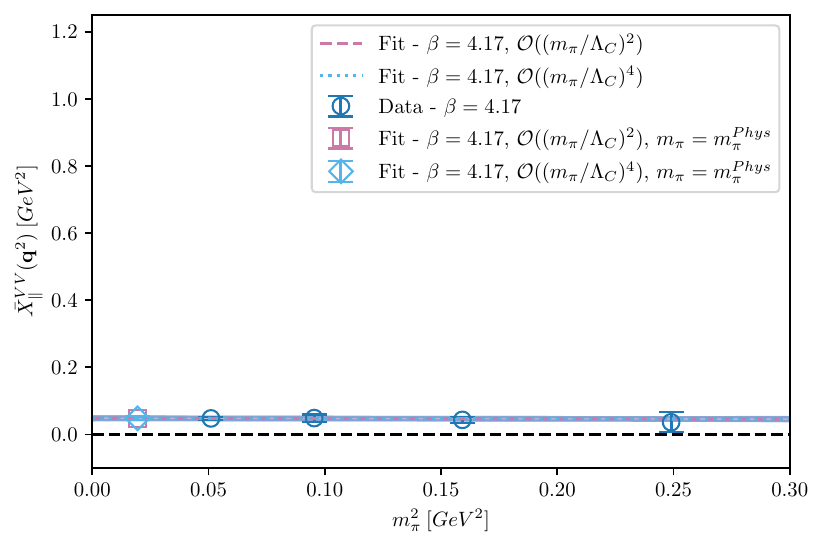}
    \end{subfigure}
    \begin{subfigure}{0.49\textwidth}
    \centering
        \includegraphics[width=\textwidth]{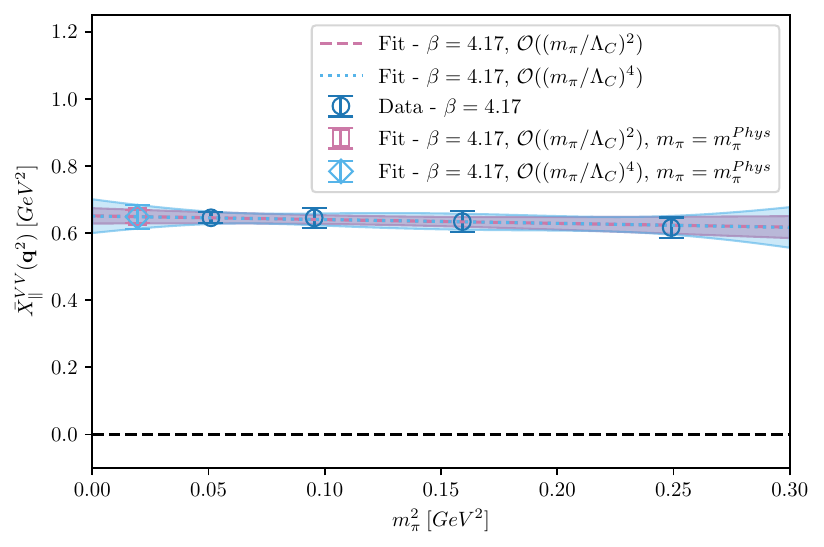}
    \end{subfigure}

    \begin{subfigure}{0.49\textwidth}
        \includegraphics[width=\textwidth]{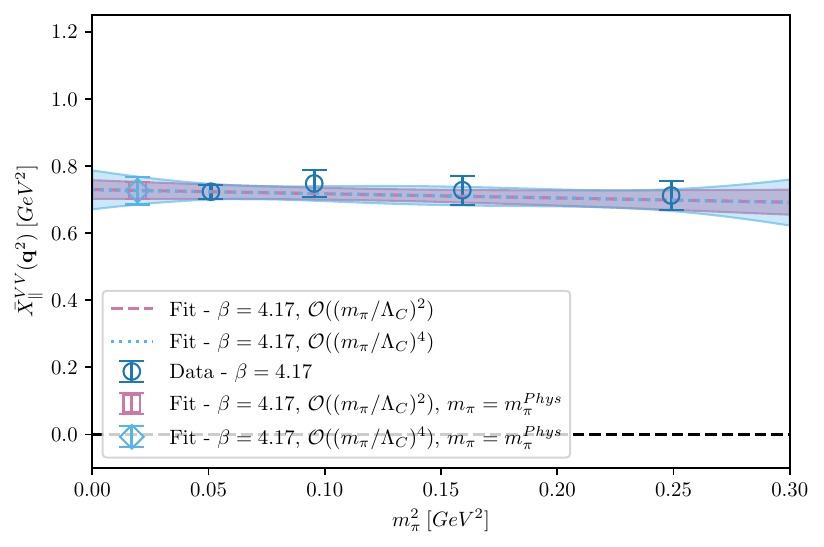}
    \end{subfigure}
    \begin{subfigure}{0.49\textwidth}
        \includegraphics[width=\textwidth]{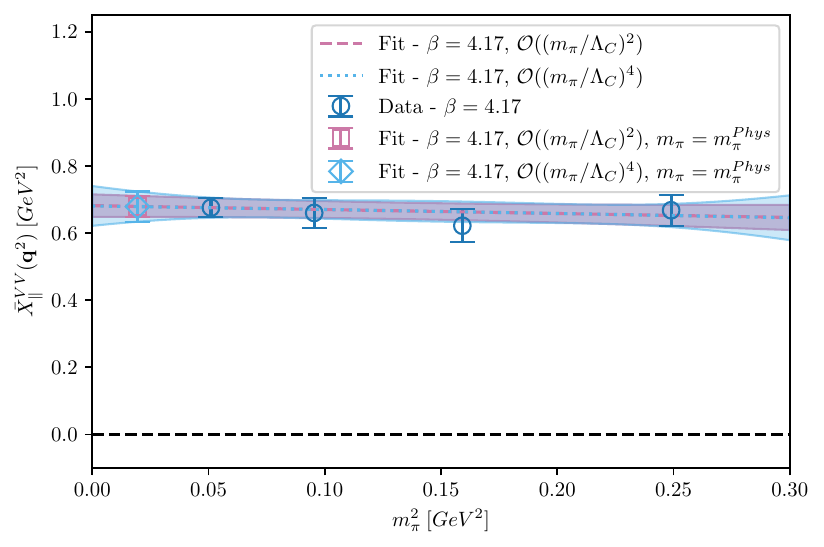}
    \end{subfigure}
    \caption{Pion-mass dependence for $\bar{X}_\parallel^{VV}(\bm{q}^2)$ at $\beta=4.17$ for each value of $\bm{q}$. The inserted momentum is from $\bm{q} = (0,0,0)$ (top left) to $\bm{q} = (1,1,1)$ (bottom right). Blue circles represent the data points for different ensembles. The data point for the ensemble ''C-$ud$2-$s$a-L'' is obtained from the fit interpolated to the same physical value of $\bm{q}$ as the other ensembles. The magenta squares and blue diamonds are the extrapolated values from the fit including terms up to $\mathcal{O}((m_\pi/\Lambda_C)^2)$ and $\mathcal{O}((m_\pi/\Lambda_C)^4)$, respectively.
    }
  	\label{fig:PionMassDependence}
\end{figure}

The extrapolation in the pion mass is shown in Fig.~\ref{fig:PionMassDependence} for different choices of $\bm{q}$. We observe that the dependence is weak and the fit results with two ansatz, {\it i.e.} with and without the $m_\pi^4$ term, are consistent with each other. The error in the physical pion mass is enhanced with the $m_\pi^4$ term because of an extra fit parameter.

\begin{figure}[tbp]
    \centering
    \begin{subfigure}{0.49\textwidth}
        \centering
        \includegraphics[width=\textwidth]{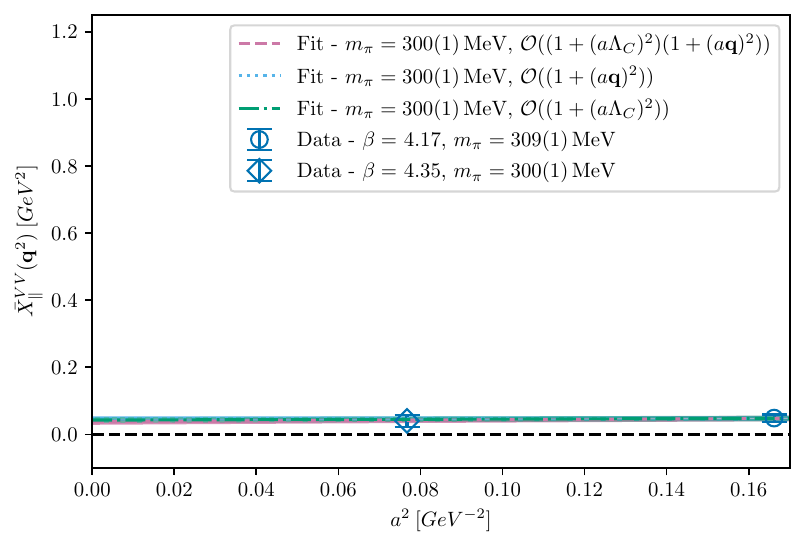}
    \end{subfigure}
    \begin{subfigure}{0.49\textwidth}
    \centering
        \includegraphics[width=\textwidth]{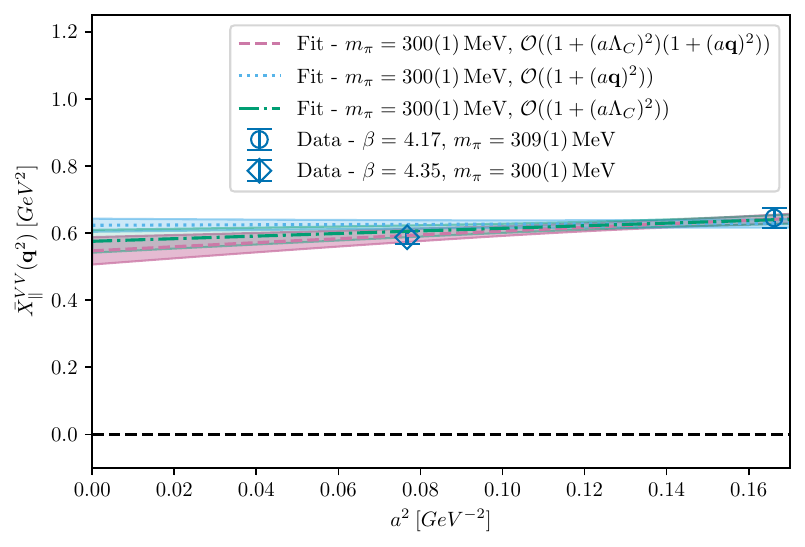}
    \end{subfigure}

    \begin{subfigure}{0.49\textwidth}
        \includegraphics[width=\textwidth]{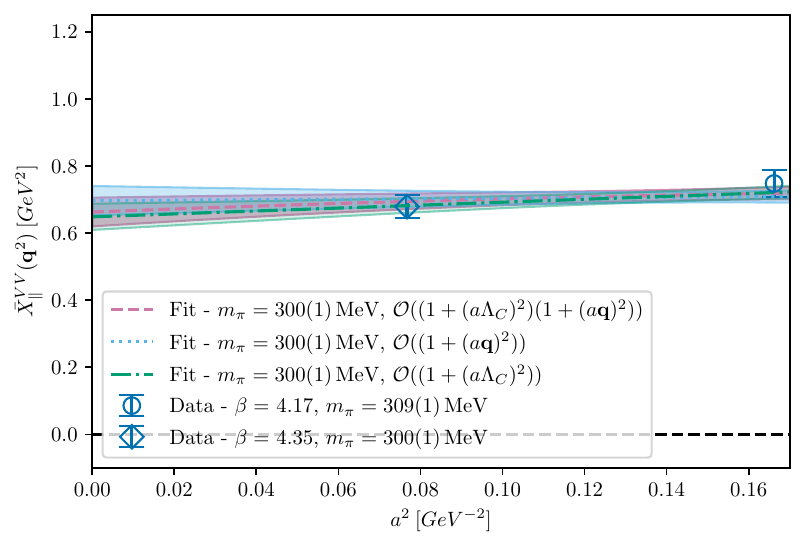}
    \end{subfigure}
    \begin{subfigure}{0.49\textwidth}
        \includegraphics[width=\textwidth]{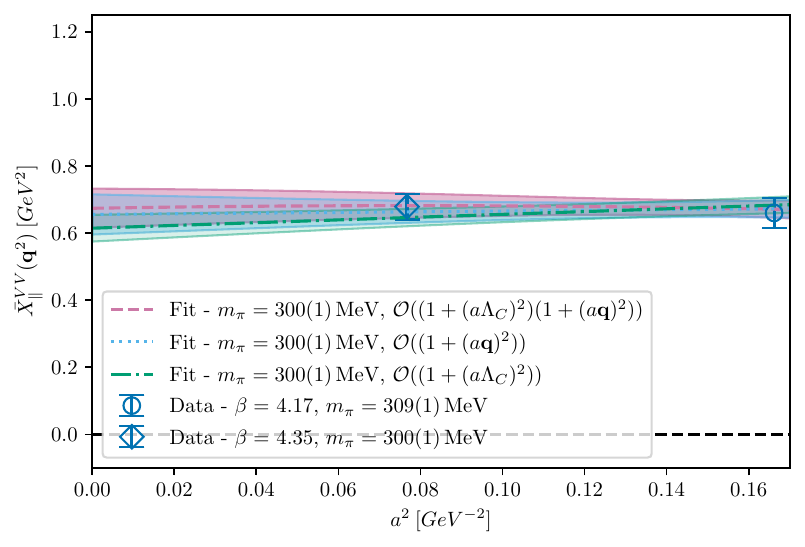}
    \end{subfigure}
    \caption{Lattice-spacing dependence of our ensembles for each value of $\bm{q}$ used in our simulations. We use the ordering starting from the top left with $\bm{q} = (0,0,0)$ and ending in the bottom right with $\bm{q} = (1,1,1)$. We show the data points which correspond to roughly the same physical pion mass in our simulations, $\beta=4.35$ with $m_\pi=\SI{300(1)}{\mega\electronvolt}$ and $\beta=4.17$ with $m_\pi=\SI{309(1)}{\mega\electronvolt}$ as orange and blue circles, respectively. The fits are obtained using different assumptions for the $a^2$ dependence. We describe cut-off effects with a combination of $\mathcal{O}(a^2)$ and $\mathcal{O}((a\bm{q})^2)$ effects (dashed line), or only one of the two (dash-dotted and dotted lines, respectively.)}
  	\label{fig:LSpacingDependence}
\end{figure}

The continuum extrapolation is shown in Fig.~\ref{fig:LSpacingDependence}. Again, the dependence on $a^2$ is weak, but we observe a small downward trend towards the continuum limit.
As the current extrapolation is based on only two data points, adding an additional data point at a different lattice spacing would yield a better quantification of the behavior in the continuum limit.

\begin{figure}[tbp]
    \centering
    \includegraphics[width=.8\textwidth]{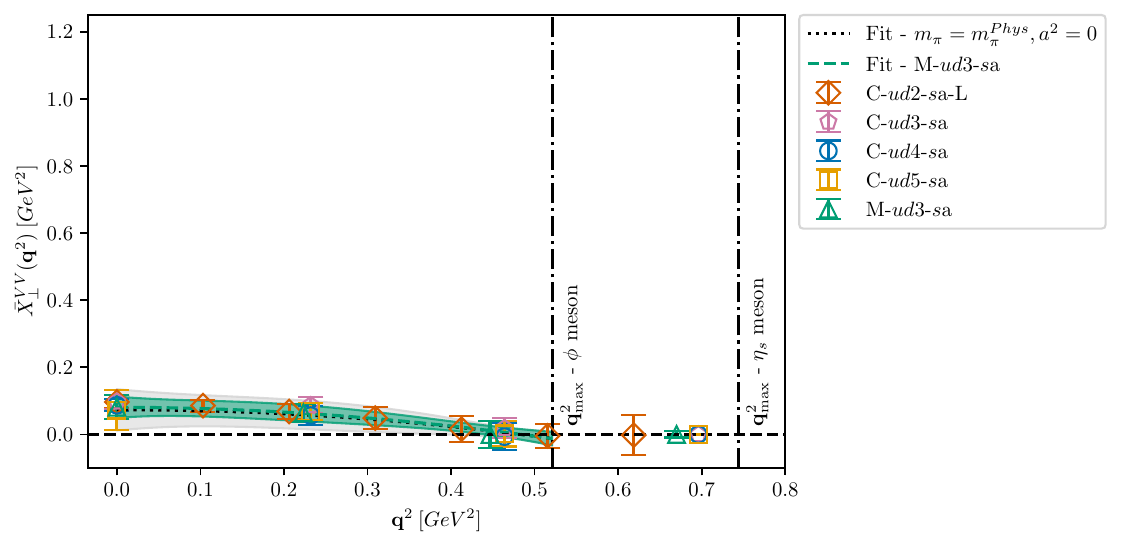}

    \centering
    \includegraphics[width=.8\textwidth]{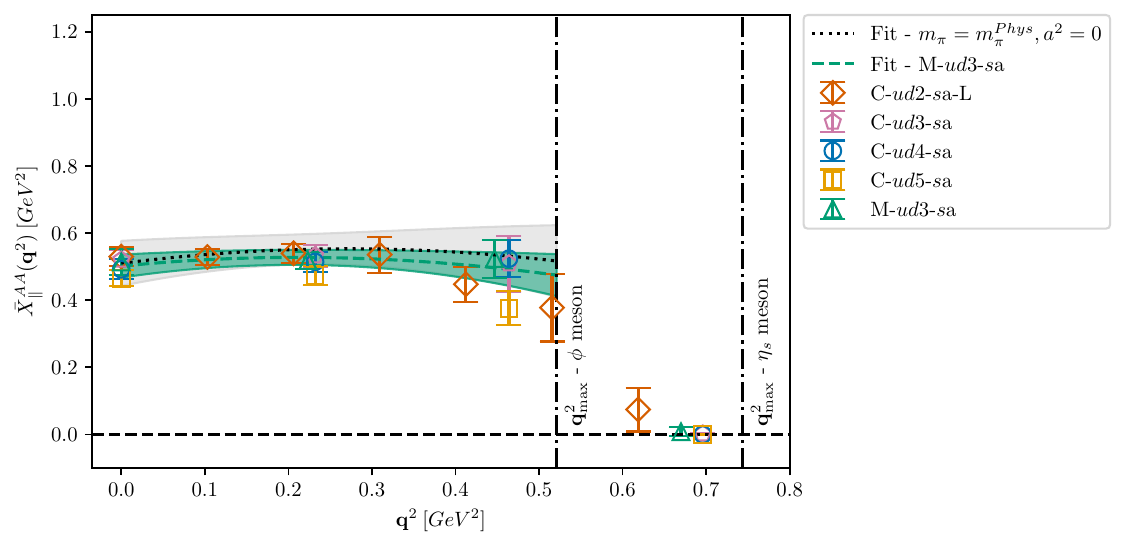}

    \centering
    \includegraphics[width=.8\textwidth]{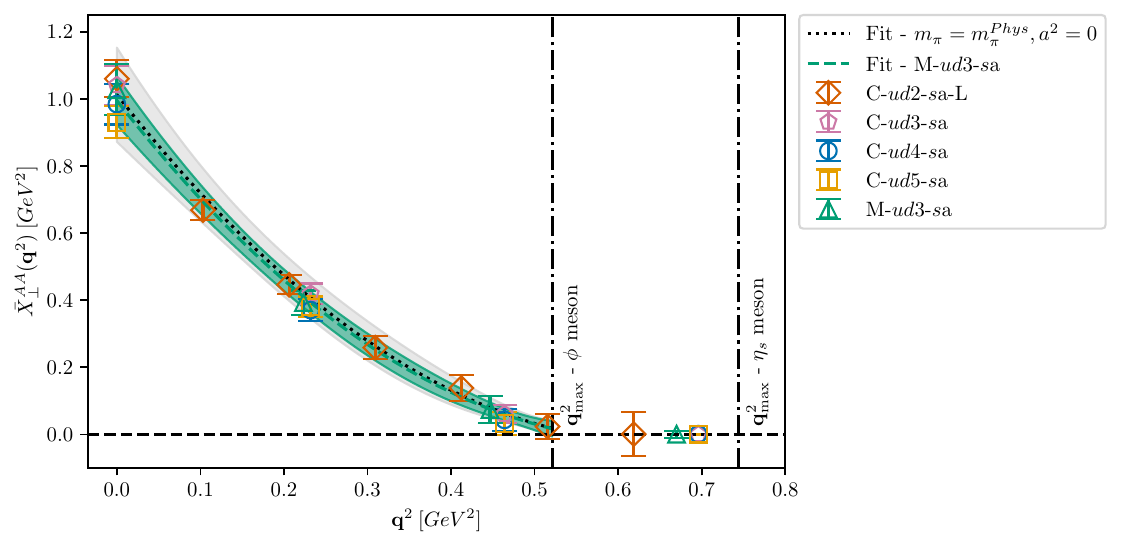}

    \caption{Same as Fig. \ref{fig:XVVParallelFitComparisonPhysiclaPoint}, but for $\bar{X}_{\perp}^{VV}(\bm{q}^2)$ (top panel), $\bar{X}_{\parallel}^{AA}(\bm{q}^2)$ (middle panel) and $\bar{X}_{\perp}^{AA}(\bm{q}^2)$ (bottom panel).}
  	\label{fig:XBarFitComparisonPhysiclaPoint}
\end{figure}

The dependence of the fit on $m_\pi^2$ and $a^2$ in the physical pion mass and continuum limit is most significant in the case studied here, {\it i.e.} for the channel of $\bar{X}_\parallel^{VV}(\bm{q}^2)$, resulting in a visible shift of the central value when compared to the fit results to individual ensembles, although it is still in agreement with the data. 
The equivalent of Fig. \ref{fig:XVVParallelFitComparisonPhysiclaPoint} for the remaining channels is shown in Fig. \ref{fig:XBarFitComparisonPhysiclaPoint} on the order of $\bar{X}_\perp^{VV}(\bm{q}^2)$, $\bar{X}_\parallel^{AA}(\bm{q}^2)$ and $\bar{X}_\perp^{AA}(\bm{q}^2)$ from top to bottom. The equivalent of Figs. \ref{fig:PionMassDependence} and \ref{fig:LSpacingDependence} for these channels can be found in Figs. \ref{fig:XVVPerpPionMassDependence}--\ref{fig:XAAPerpLSpacingDependence} in the Appendix~\ref{sec:TablesAndFigures}. We find that they exhibit a milder dependence for both limits.

\section{\label{sec:InclusiveRate}Determination of the inclusive rate}

\begin{figure}[tbp]
    \centering
    \includegraphics[width=\textwidth]{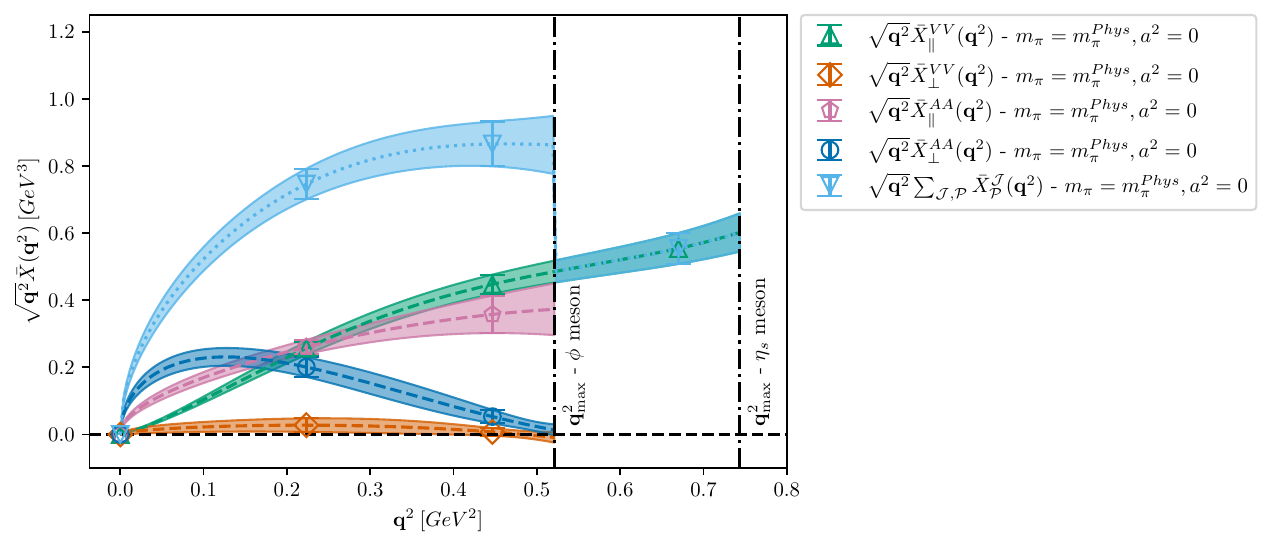}
    \caption{Differential decay rate $\sqrt{\bm{q}^2} \bar{X}(\bm{q}^2)$ constituting the integrand in \eqref{equ:TotalRateInclusive}. We show the fit results to the individual channels $V_\mu^\dagger V_\mu$ and $A_\mu^\dagger A_\mu$ in the continuum and physical pion mass limits. The inverted triangles represent the sum over individual contributions.}
    \label{fig:IntegrandTotalRate}
\end{figure}

Using the parameters determined from the individual fits, summarized in Tabs. \ref{tab:XVVParallel_Multifit}, \ref{tab:XVVPerp_Multifit}, \ref{tab:XAAParallel_Multifit} and \ref{tab:XAAPerp_Multifit}, defined in the previous section, we now determine the continuum and pion-mass extrapolated contribution of each individual channel contributing to $\bar{X}(\bm{q}^2)$. We multiply the results by $\sqrt{\bm{q}^2}$ in order to construct the integrand of the total decay rate defined in \eqref{equ:TotalRateInclusive}. The results are shown in Fig.~\ref{fig:IntegrandTotalRate}, where we show the results for the individual channels and their sum. We remark that the curves shown are based solely on the fit and the data points are simply included to highlight the $\bm{q}^2$ values used in our simulations. As the lightest hadronic final state between $\bar{X}^{VV}_{\parallel}(\bm{q}^2)$ and the rest, {\it i.e.} the pseudoscalar $\eta_s$ and vector $\phi$ meson, respectively, require different values of $\bm{q}^2_{\text{max}}$, we exclude the highest $\bm{q}^2$ value for all channels except $\bar{X}^{VV}_{\parallel}(\bm{q}^2)$.
Their physical value is expected to be 0 due to kinematical constraints. This also gives rise to the discontinuity observed at $\bm{q}^2_{\text{max},\phi}$, as beyond this threshold the only contribution comes from $\bar{X}_{\parallel}^{VV}(\bm{q}^2)$.

\begin{table}[tbp]
	\centering
	\begin{tabular}{c | l }
	& $\frac{\Gamma}{|V_{cs}|^2} \times 10^{13}$ \\
	\hline
	$\bar{X}^{VV}_{\parallel}$ & $\nonsymerr{0.480(31)}{+0.044}{-0.049}$ \\
	$\bar{X}^{VV}_{\perp}$ & $\nonsymerr{0.017(12)}{+0}{-0.008}$ \\
	$\bar{X}^{AA}_{\parallel}$ & $\nonsymerr{0.248(26)}{+0.001}{-0.021}$ \\
	$\bar{X}^{AA}_{\perp}$ & $\nonsymerr{0.139(20)}{+0.002}{-0.010}$ \\
    \hline
    total & $\nonsymerr{0.884(47)}{+0.044}{-0.055}$
	\end{tabular}
	\caption{Results for the $\bm{q}^2$ integral for $\Gamma/|V_{cs}|^2 = G_F^2/(24\pi^3)\int_{0}^{\bm{q}^2_{\text{max}}} d\bm{q}^2 \sqrt{\bm{q}^2} \bar{X}(\bm{q}^2)$, in $\text{GeV}$, for individual channels. The first error is the statistical error from the fit to the data, while the second error is the systematic error associated with the continuum and physical pion mass extrapolation.}
	\label{tab:XBarIntegral}
\end{table}

This also motivates us to perform the $\bm{q}^2$-integration for each contribution individually, since we would need to take the discontinuity into account otherwise. We perform the integration analytically based on the fit prescription \eqref{equ:ExtrapolationFit}, and the total decay rate is then constructed as the sum of the individual contributions. The integration results for the individual channels are given in Tab. \ref{tab:XBarIntegral}. 

For the inclusive decay rate, we obtain $\Gamma/|V_{cs}|^2 = \nonsymerr{0.884(47)}{+0.044}{-0.055}\times10^{-13} \, \si{\giga\electronvolt}$, where the first error is  statistical  from our fit to the data, combining the statistical noise of the data and our estimates for the $\sigma\to 0$ and $V\to\infty$ limits. The second error is the systematic error associated with the continuum and physical pion mass extrapolation, defined as the largest spread in central values between the different fit variations. Improvements for both uncertainties can be expected once higher statistics become available and additional ensembles are included, enabling more controlled chiral and continuum extrapolations. Nonetheless, this calculation constitutes a determination of the inclusive decay rate taking all systematic uncertainties from a lattice calculation into account.
This number is in good agreement with our previous findings $\Gamma/|V_{cs}|^2 = \SI{0.886(27)e-13}{\giga\electronvolt}$ (single lattice spacing and sea-pion mass), where our new determination has a larger error of around 8\%. It is also in good agreement with the result presented in \cite{DeSantis:2025qbb}, $\Gamma/|V_{cs}|^2 = 0.853(46)(30)\times 10^{-13} \, \si{\giga\electronvolt}$, which was determined using the HLT reconstruction approach. We remark that our results do not include contributions from disconnected diagrams, which have been analyzed in \cite{DeSantis:2025qbb}, and final states including $\eta$ and $\eta'$ may affect the final results. An additional point of concern is that the $\phi$ meson is stable for our simulation parameters, and the finite-volume effects, as well as the chiral extrapolation, may behave differently once the $\phi\to K\bar{K}$ decays are possible. With these caveats in mind, we compare with the experimental data from the BESIII collaboration \cite{BESIII:2021duu}, $\Gamma=\SI{0.827(22)e-13}{\giga\electronvolt}$, and CLEO collaboration \cite{CLEO:2009uah}, $\Gamma=\SI{0.856(55)e-13}{\giga\electronvolt}$. This yields the following values for $|V_{cs}|$
\begin{align}
    |V_{cs}|_{\text{BESIII}} &= \nonsymerr{0.967(28)}{+0.024}{-0.029} \, , \\ \nonumber
    |V_{cs}|_{\text{CLEO}} &= \nonsymerr{0.984(41)}{+0.024}{-0.032} \, ,
\end{align}
where the first error is the combined statistical error from theory and experiment, and the second error is the systematic error for the chiral and continuum extrapolations. They are comparable to the PDG value $|V_{cs}| = 0.975(6)$ \cite{ParticleDataGroup:2024cfk}, albeit with a larger error budget on our lattice QCD determination. We note that the analysis performed in this paper does not take into account the contributions of the Cabibbo-suppressed $c\to d$ and $u\to s$ transitions to the inclusive rate, which is a subject of future investigations. 

\section{\label{sec:Conclusion}Conclusion and Outlook}

We performed a lattice-QCD analysis of the inclusive semileptonic decay of the $D_s$ meson. We build on our previous work \cite{Kellermann:2025pzt} using the Chebyshev-polynomial approximation for the energy integral, concentrating in this work on the chiral and continuum extrapolations. We employ a global fitting strategy parameterizing the $\bm{q}^2$, $m_\pi^2$ and $a^2$ dependencies simultaneously. This constitutes another step towards a fully controlled lattice prediction for the inclusive decay rate, although some caveats still remain. For one, the current results are based on only two lattice spacings for the continuum extrapolation, and we plan to include an ensemble with a third lattice spacing in the future for a more robust extrapolation. Additionally, our current approach still relies on a model of two-body states combined with lattice data to estimate the infinite-volume extrapolation, see Ref. \cite{Kellermann:2025pzt}. While our model predicts only minor corrections from finite-volume effects, a more comprehensive approach using numerical data from multiple volumes is a target for future studies. Additionally, with our setup the $K\bar{K}$ states are always heavier than the $\phi$, since the $u/d$ quark mass is heavier than physical. A study under a setup where the $\phi$ meson can decay to $K\bar{K}$ could provide further insights to the size of potential finite--volume effects, as well as effects from the light quark masses.

The work on the $B_{(s)}$ decay, along the direction of \cite{Barone:2023tbl}, is also underway. Additionally, a project using the four-point correlation functions to study the processes involving $P$-wave final states, as attempted in \cite{Hu:2025hpn}, is also underway.

Furthermore, the extension of the analysis to other observables, such as lepton-energy and $q^2$ moments, as was done in \cite{DeSantis:2025qbb}, is also planned, which will increase the pool of observables that can be compared to experimental observations or OPE-based calculations, see {\it e.g.} \cite{Gambino:2022dvu}. We are also studying ways for comparing experiment and theory at finite smearing width, which would render the $\sigma\to 0$ limit unnecessary~\cite{Juttner:2026fui}.

\begin{acknowledgments}
The numerical calculations of the JLQCD collaboration were performed on SX-Aurora TSUBASA at the High Energy Accelerator Research Organization (KEK) under its Particle, Nuclear and Astrophysics Simulation Program, as well as on Fugaku through the HPCI System Research Project (Project ID: hp220056). The works of S.H., R.K. and T.K. are supported in part by JSPS KAKENHI Grant Numbers 22H00138, 22K21347, 23K20846 and 25K01007, and by the Post-K and Fugaku supercomputer project through the Joint Institute for Computational Fundamental Science (JICFuS). T.K. is also supported by the U.S.-Japan Science and Technology Cooperation Program in High Energy Physics (Project ID: 2024-40-2). A.J. is supported by the Eric \& Wendy Schmidt Fund for Strategic Innovation (grant agreement SIF-2023-004).
\end{acknowledgments}

\clearpage
\appendix

\section{\label{sec:TablesAndFigures}Tables and Figures}

\begin{figure}[tbp]
    \centering
    \includegraphics[width=.6\textwidth]{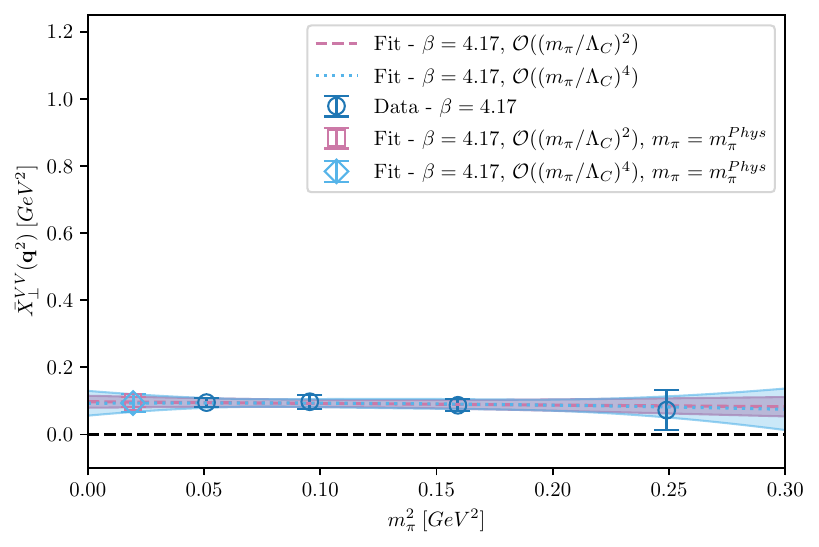}

    \centering
    \includegraphics[width=.6\textwidth]{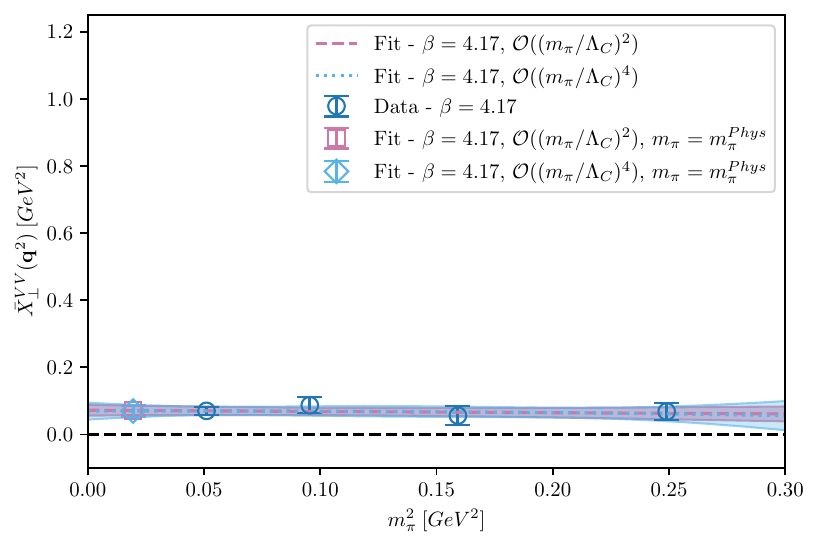}

    \centering
    \includegraphics[width=.6\textwidth]{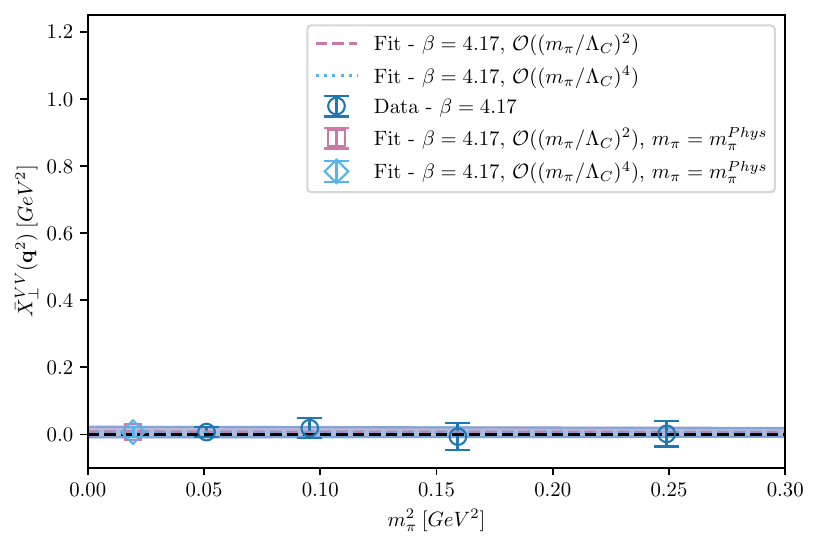}

    \caption{Same as Fig. \ref{fig:PionMassDependence}, but for $\bar{X}_{\perp}^{VV}(\bm{q}^2)$. The plot on the top has $\bm{q} = (0,0,0)$, going up to $\bm{q} = (0,1,1)$ for the bottom plot.}
  	\label{fig:XVVPerpPionMassDependence}
\end{figure}

\begin{figure}[tbp]
    \centering
    \includegraphics[width=.6\textwidth]{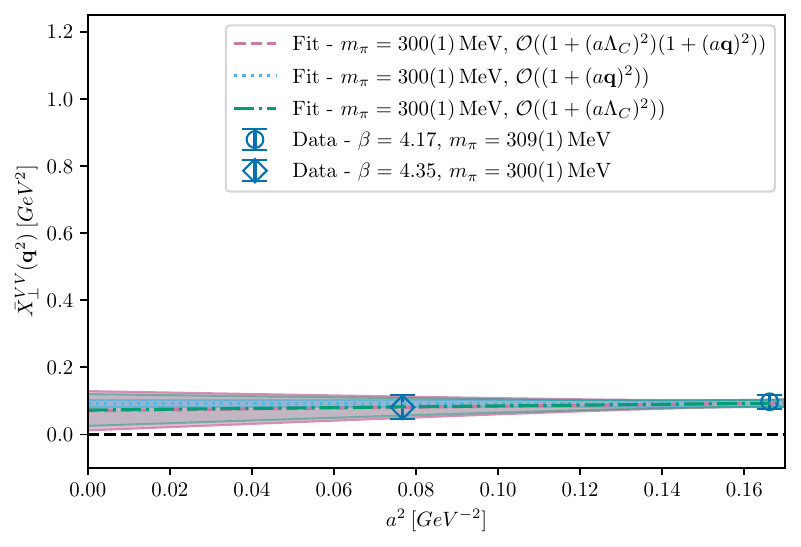}

    \centering
    \includegraphics[width=.6\textwidth]{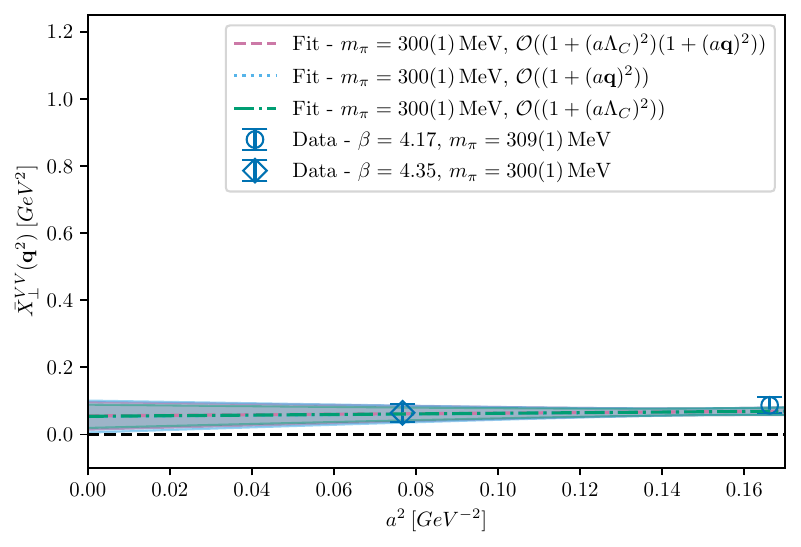}

    \centering
    \includegraphics[width=.6\textwidth]{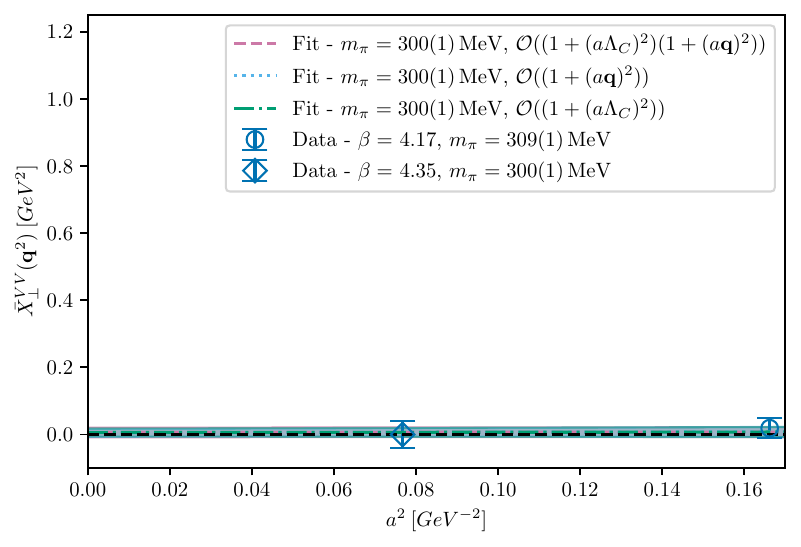}

    \caption{Same as Fig. \ref{fig:LSpacingDependence}, but for $\bar{X}_{\perp}^{VV}(\bm{q}^2)$. The plot on the top has $\bm{q} = (0,0,0)$, going up to $\bm{q} = (0,1,1)$ for the bottom plot.}
  	\label{fig:XVVPerpLSpacingDependence}
\end{figure}

\begin{figure}[tbp]
    \centering
    \includegraphics[width=.6\textwidth]{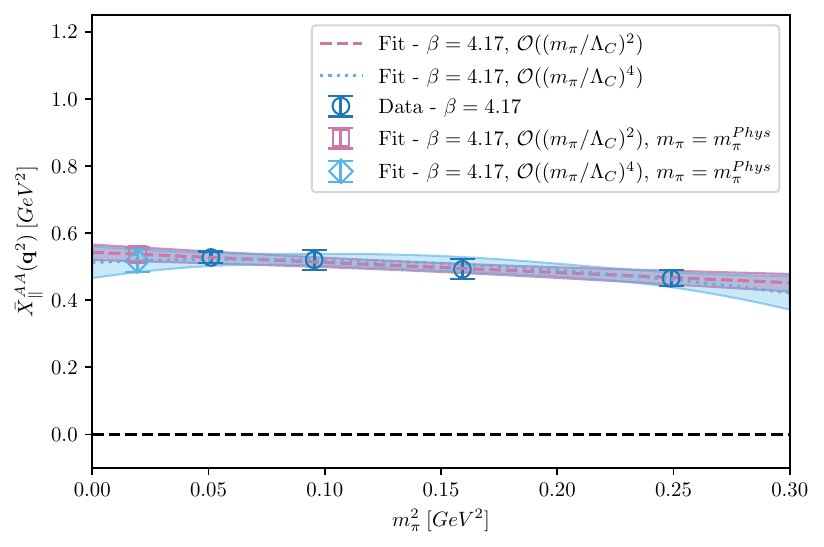}

    \centering
    \includegraphics[width=.6\textwidth]{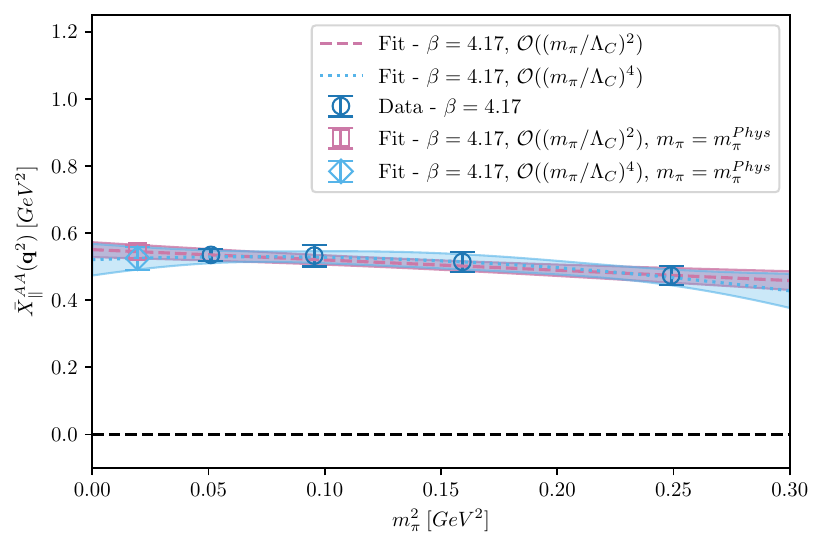}

    \centering
    \includegraphics[width=.6\textwidth]{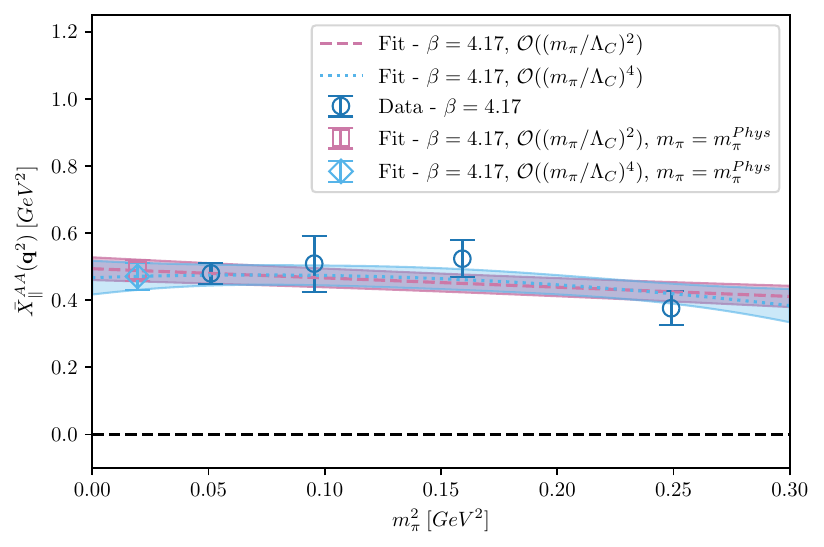}

    \caption{Same as Fig. \ref{fig:XVVPerpPionMassDependence}, but for $\bar{X}_{\parallel}^{AA}(\bm{q}^2)$.}
  	\label{fig:XAAParallelPionMassDependence}
\end{figure}

\begin{figure}[tbp]
    \centering
    \includegraphics[width=.6\textwidth]{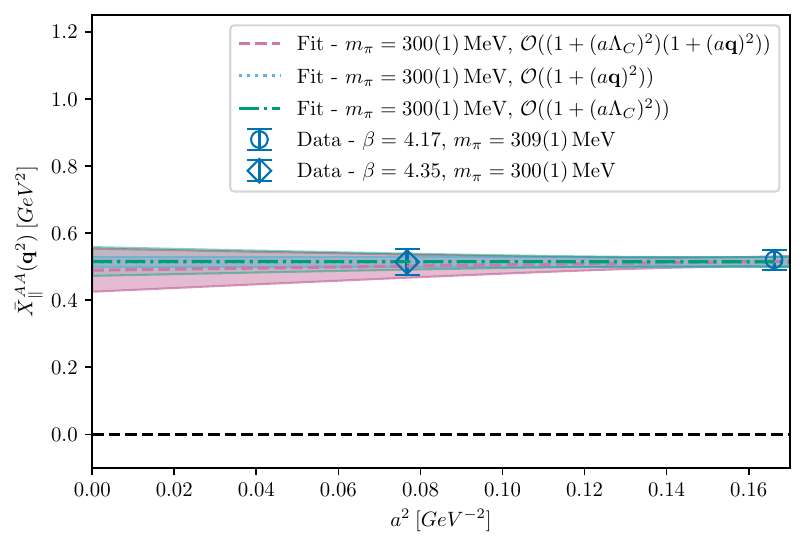}

    \centering
    \includegraphics[width=.6\textwidth]{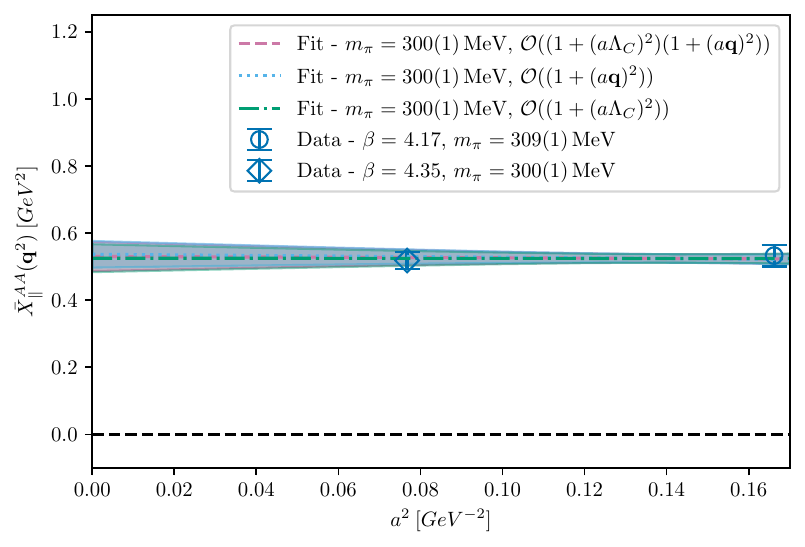}

    \centering
    \includegraphics[width=.6\textwidth]{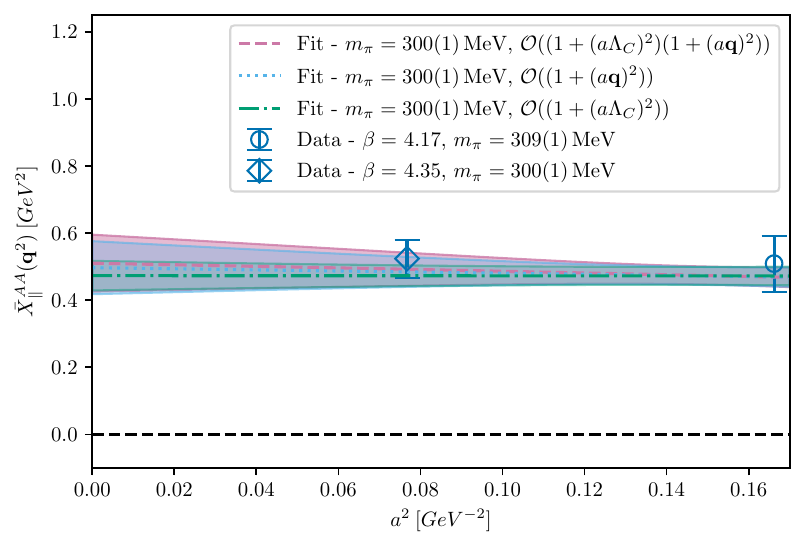}

    \caption{Same as Fig. \ref{fig:XVVPerpLSpacingDependence}, but for $\bar{X}_{\parallel}^{AA}(\bm{q}^2)$.}
  	\label{fig:XAAParallelLSpacingDependence}
\end{figure}

\begin{figure}[tbp]
    \centering
    \includegraphics[width=.6\textwidth]{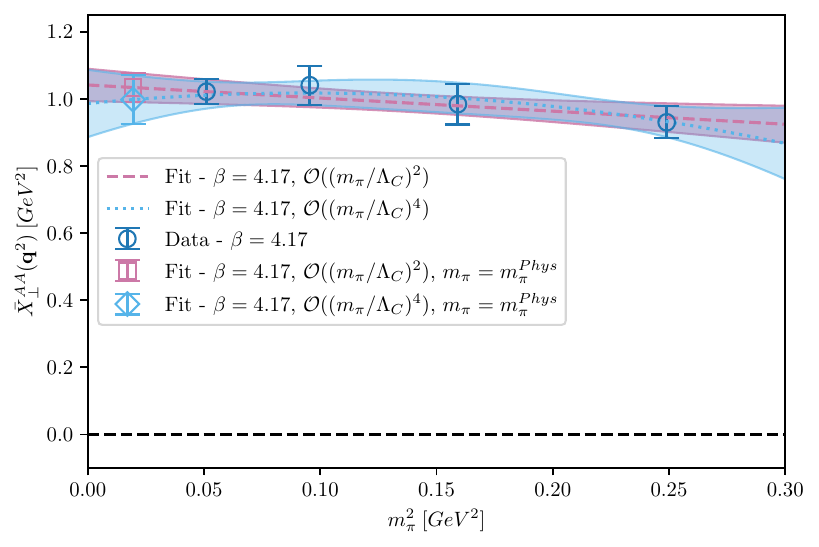}

    \centering
    \includegraphics[width=.6\textwidth]{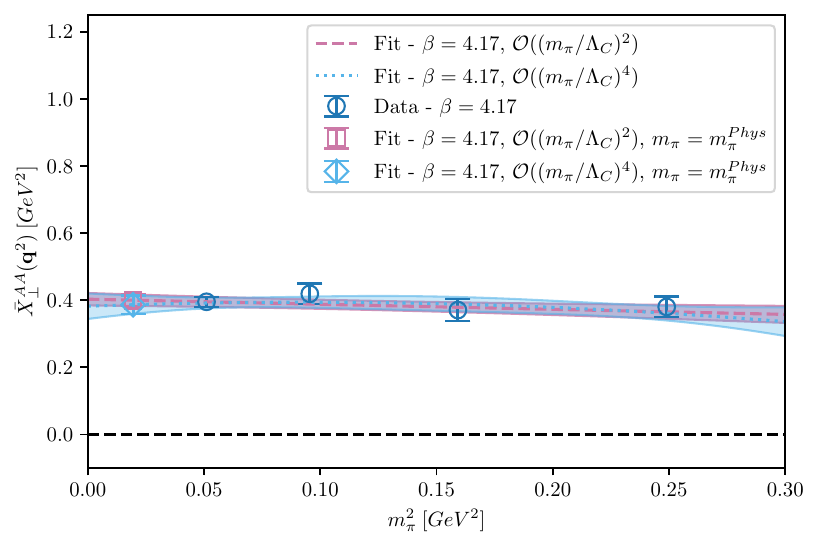}

    \centering
    \includegraphics[width=.6\textwidth]{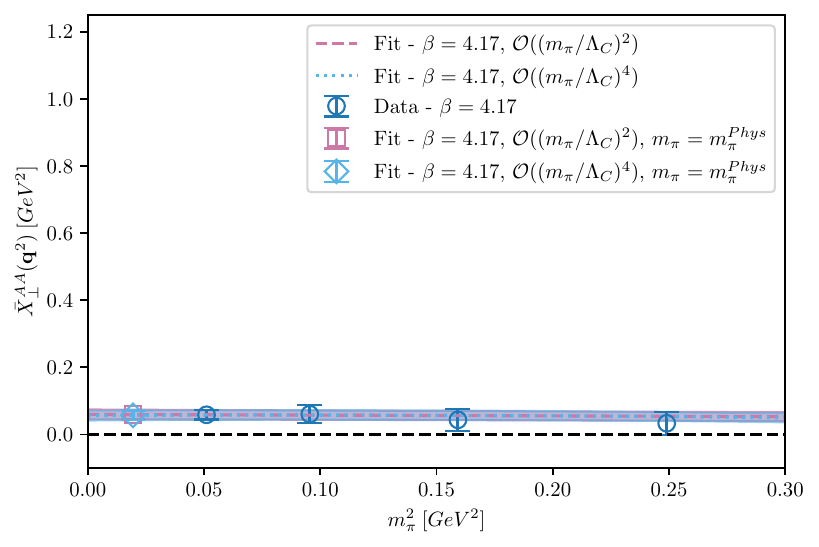}

    \caption{Same as Fig. \ref{fig:XVVPerpPionMassDependence}, but for $\bar{X}_{\perp}^{AA}(\bm{q}^2)$.}
  	\label{fig:XAAPerpPionMassDependence}
\end{figure}

\begin{figure}[tbp]
    \centering
    \includegraphics[width=.6\textwidth]{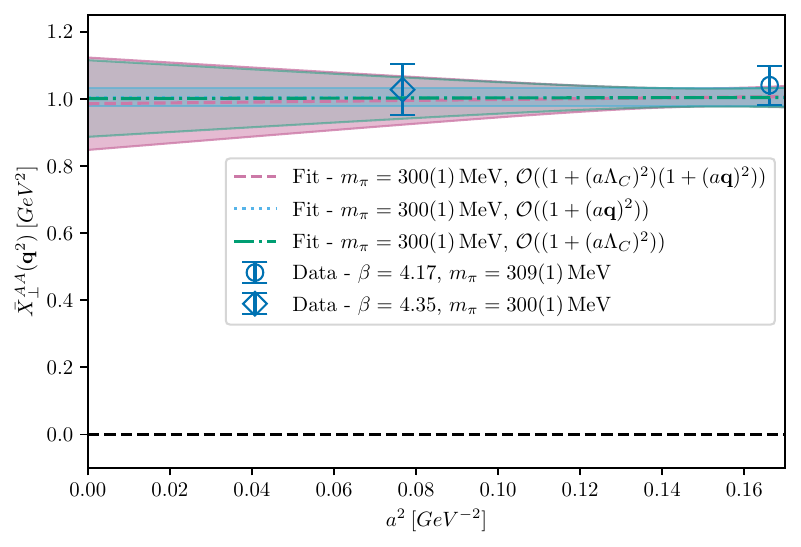}

    \centering
    \includegraphics[width=.6\textwidth]{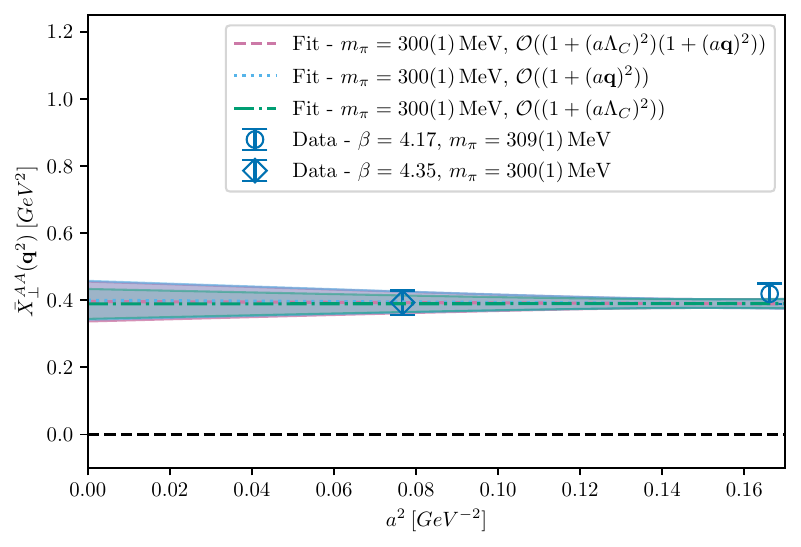}

    \centering
    \includegraphics[width=.6\textwidth]{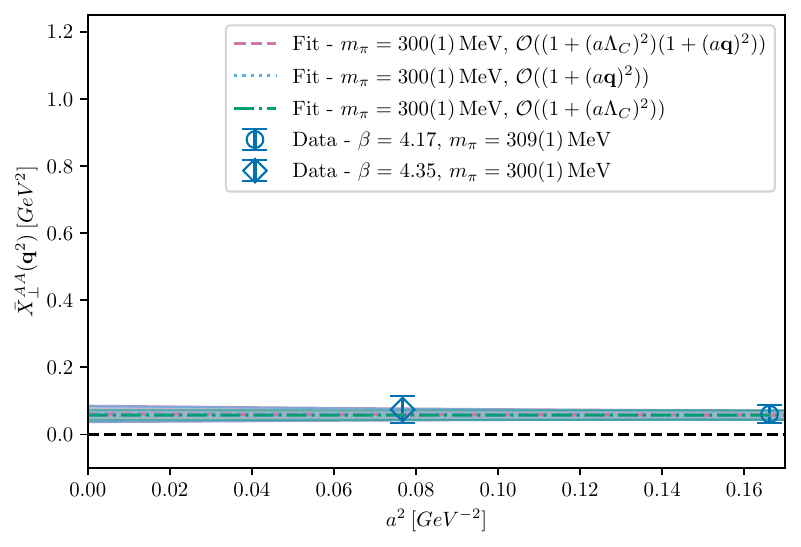}

    \caption{Same as Fig. \ref{fig:XVVPerpLSpacingDependence}, but for $\bar{X}_{\perp}^{AA}(\bm{q}^2)$.}
  	\label{fig:XAAPerpLSpacingDependence}
\end{figure}

We summarize the input data and fit parameters in order of $\bar{X}^{VV}_{\parallel}(\bm{q}^2)$, $\bar{X}^{VV}_{\perp}(\bm{q}^2)$, $\bar{X}^{AA}_{\parallel}(\bm{q}^2)$ and $\bar{X}^{AA}_{\perp}(\bm{q}^2)$ in Tabs. \ref{tab:XVVParallelDataForFit} -- \ref{tab:XAAPerp_Multifit}. We further show the pion mass and lattice spacing dependence in Figs. \ref{fig:XVVPerpPionMassDependence}--\ref{fig:XAAPerpLSpacingDependence} for all channels except $\bar{X}^{VV}_{\parallel}(\bm{q}^2)$, which has been discussed in the main body.

For all channels except $\bar{X}^{VV}_{\parallel}(\bm{q}^2)$ the lightest hadronic final state is the vector $\phi$ meson, and hence the highest momentum in our simulations $\bm{q} = (1,1,1)$ (or $(1,1,2)$ for C-$ud3$-$s$a-L) is already above the kinematical threshold. The data points are shown in the plots, but are excluded in the fits and hence not included in the tables. Consequently, the polynomial order of the $\bm{q}^2$ dependence in the fit prescription \eqref{equ:ExtrapolationFit} is chosen as $N_p = 2$, reducing the number of fit parameters from seven to six.

\begin{table}[tbp]
  \centering
  \begin{tabular}{l || r r r r r r r}
    & \multicolumn{7}{c}{$\bar{X}^{VV}_{\parallel}(\bm{q}^2) \, [\text{GeV}^2]$} \\\cline{2-8}
    ID & \multicolumn{1}{c}{(0,0,0)} & \multicolumn{1}{c}{(0,0,1)} & \multicolumn{1}{c}{(0,1,1)} & \multicolumn{1}{c}{(1,1,1)} & \multicolumn{1}{c}{(0,0,2)} & \multicolumn{1}{c}{(0,1,2)} & \multicolumn{1}{c}{(1,1,2)} \\
  \hline
    C-$ud3$-$s$a & $\num{0.049(11)}$ & $\num{0.645(29)}$ & $\num{0.748(40)}$ & $\num{0.660(45)}$ & & & \\
    C-$ud4$-$s$a & $\num{0.043(9)}$ & $\num{0.634(31)}$ & $\num{0.728(44)}$ & $\num{0.622(49)}$ & & & \\
    C-$ud5$-$s$a & $\num{0.036(29)}$ & $\num{0.617(30)}$ & $\num{0.712(43)}$ & $\num{0.668(46)}$ & & & \\
  \hline
  \hline
  C-$ud2$-$s$a-L & $\num{0.048(10)}$ & $\num{0.422(15)}$ & $\num{0.595(22)}$ & $\num{0.688(28)}$ & $\num{0.725(34)}$ & $\num{0.720(39)}$ & $\num{0.713(42)}$ \\
  \hline
  \hline
  M-$ud3$-$s$a & $\num{0.040(18)}$ & $\num{0.588(20)}$ & $\num{0.679(33)}$ & $\num{0.679(38)}$ & & & \\
  \hline
  \end{tabular}
  \caption{Values of $\bar{X}^{VV}_{\parallel}(\bm{q}^2)$ for all values of $\bm{q}^2$ determined for different ensembles entering the fit to perform the chiral and continuum extrapolations. The values contain corrections from the model--based infinite volume corrections and the $\sigma\to 0$ extrapolation.
  }
  \label{tab:XVVParallelDataForFit}
\end{table}
\begin{table}[tbp]
	\centering
	\begin{tabular}{r r r r r r r}
        \multicolumn{1}{c}{$c_{0}\, [\text{GeV}^2]$} & \multicolumn{1}{c}{$c_{1}\,[-]$} & \multicolumn{1}{c}{$c_{2}\,[\text{GeV}^{-2}]$} & \multicolumn{1}{c}{$c_{3}\,[\text{GeV}^{-4}]$} & \multicolumn{1}{c}{$\tilde{c}_{m_\pi}\,[-]$} & \multicolumn{1}{c}{$\tilde{c}_{a}\,[-]$} & \multicolumn{1}{c}{$\tilde{c}_{a\bm{q}}\,[-]$} \\
        \hline
	\multicolumn{1}{c}{$\num{0.04(1)}$} & \multicolumn{1}{c}{$\num{3.52(34)}$} & \multicolumn{1}{c}{$\num{-6.45(1.03)}$} & \multicolumn{1}{c}{$\num{3.93(81)}$} & \multicolumn{1}{c}{$\num{-0.08(12)}$} & \multicolumn{1}{c}{$\num{3.19(1.71)}$} & \multicolumn{1}{c}{$\num{-1.81(1.03)}$} \\
	\hline
        \hline
       $\num{1.00}$ & $\num{0.59}$ & $\num{-0.52}$ & $\num{0.42}$ & $\num{-0.12}$ & $\num{-0.68}$ & $\num{0.55}$ \\
$\num{0.59}$ & $\num{1.00}$ & $\num{-0.94}$ & $\num{0.85}$ & $\num{-0.19}$ & $\num{-0.91}$ & $\num{0.70}$ \\
$\num{-0.52}$ & $\num{-0.94}$ & $\num{1.00}$ & $\num{-0.97}$ & $\num{0.15}$ & $\num{0.82}$ & $\num{-0.78}$ \\
$\num{0.42}$ & $\num{0.85}$ & $\num{-0.97}$ & $\num{1.00}$ & $\num{-0.13}$ & $\num{-0.69}$ & $\num{0.67}$ \\
$\num{-0.12}$ & $\num{-0.19}$ & $\num{0.15}$ & $\num{-0.13}$ & $\num{1.00}$ & $\num{-0.08}$ & $\num{-0.07}$ \\
$\num{-0.68}$ & $\num{-0.91}$ & $\num{0.82}$ & $\num{-0.69}$ & $\num{-0.08}$ & $\num{1.00}$ & $\num{-0.75}$ \\
$\num{0.55}$ & $\num{0.70}$ & $\num{-0.78}$ & $\num{0.67}$ & $\num{-0.07}$ & $\num{-0.75}$ & $\num{1.00}$ \\
    \hline
	\end{tabular}
    \caption{Fit results for $\bar{X}_{\parallel}^{VV}(\bm{q}^2)$ using all available ensembles. The first line shows the fit parameters determined from the fit while the second line and beyond contain the correlation matrix of the fit parameters. The $\chi^2/\text{d.o.f.}$ from the correlated fit is given by $0.48$.}
	\label{tab:XVVParallel_Multifit}
\end{table}

\begin{table}[tbp]
  \centering
  \begin{tabular}{l || r r r r r r}
    & \multicolumn{6}{c}{$\bar{X}^{VV}_{\perp}(\bm{q}^2) \, [\text{GeV}^2]$} \\\cline{2-7}
    ID & \multicolumn{1}{c}{(0,0,0)} & \multicolumn{1}{c}{(0,0,1)} & \multicolumn{1}{c}{(0,1,1)} & \multicolumn{1}{c}{(1,1,1)} & \multicolumn{1}{c}{(0,0,2)} & \multicolumn{1}{c}{(0,1,2)} \\
  \hline
    C-$ud3$-$s$a & $\num{0.097(21)}$ & $\num{0.087(24)}$ & $\num{0.019(30)}$ & & & \\
    C-$ud4$-$s$a & $\num{0.087(18)}$ & $\num{0.056(29)}$ & $\num{-0.006(40)}$ & & & \\
    C-$ud5$-$s$a & $\num{0.072(59)}$ & $\num{0.068(25)}$ & $\num{0.002(38)}$ & & & \\
  \hline
  \hline
  C-$ud2$-$s$a-L & $\num{0.097(19)}$ & $\num{0.086(18)}$ & $\num{0.068(24)}$ & $\num{0.049(32)}$ & $\num{0.016(38)}$ & $\num{-0.004(36)}$ \\  
  \hline
  \hline
  M-$ud3$-$s$a & $\num{0.081(36)}$ & $\num{0.064(26)}$ & $\num{-0.000(40)}$ & & & \\
  \hline
  \end{tabular}
  \caption{Same as \ref{tab:XVVParallelDataForFit} but with the data entering the fit for $\bar{X}^{VV}_{\perp}(\bm{q}^2)$.}
  \label{tab:XVVPerpDataForFit}
\end{table}
\begin{table}[tbp]
	\centering
	\begin{tabular}{r r r r r r}
        \multicolumn{1}{c}{$c_{0} \, [\text{GeV}^2]$} & \multicolumn{1}{c}{$c_{1} \, [-]$} & \multicolumn{1}{c}{$c_{2} \, [\text{GeV}^{-2}]$} & \multicolumn{1}{c}{$\tilde{c}_{m_\pi} \, [-]$} & \multicolumn{1}{c}{$\tilde{c}_{a} \, [-]$} & \multicolumn{1}{c}{$\tilde{c}_{a\bm{q}} \, [-]$} \\
        \hline
	\multicolumn{1}{c}{$\num{0.07(6)}$} & \multicolumn{1}{c}{$\num{0.01(41)}$} & \multicolumn{1}{c}{$\num{-0.33(72)}$} & \multicolumn{1}{c}{$\num{-0.25(65)}$} & \multicolumn{1}{c}{$\num{3.90(1355)}$} & \multicolumn{1}{c}{$\num{-1.46(2202)}$} \\
	\hline
        \hline
        $\num{1.00}$ & $\num{-0.62}$ & $\num{0.35}$ & $\num{-0.22}$ & $\num{-0.98}$ & $\num{0.61}$ \\
$\num{-0.62}$ & $\num{1.00}$ & $\num{-0.95}$ & $\num{0.15}$ & $\num{0.58}$ & $\num{-0.97}$ \\
$\num{0.35}$ & $\num{-0.95}$ & $\num{1.00}$ & $\num{-0.10}$ & $\num{-0.31}$ & $\num{0.91}$ \\
$\num{-0.22}$ & $\num{0.15}$ & $\num{-0.10}$ & $\num{1.00}$ & $\num{0.05}$ & $\num{-0.15}$ \\
$\num{-0.98}$ & $\num{0.58}$ & $\num{-0.31}$ & $\num{0.05}$ & $\num{1.00}$ & $\num{-0.59}$ \\
$\num{0.61}$ & $\num{-0.97}$ & $\num{0.91}$ & $\num{-0.15}$ & $\num{-0.59}$ & $\num{1.00}$ \\
        \hline
	\end{tabular}
    \caption{Same as Tab. \ref{tab:XVVParallel_Multifit} but for $\bar{X}_{\perp}^{VV}(\bm{q}^2)$. Since the lightest state for this channel is the vector $\phi$--meson, the highest $\bm{q}^2$ value lies above the kinematical threshold, and we hence use a different polynomial to parametrize the $\bm{q}^2$--dependence, the number of fit parameters is reduced by one. The $\chi^2/\text{d.o.f.}$ from the correlated fit is given by $0.12$.}
	\label{tab:XVVPerp_Multifit}
\end{table}

\begin{table}[tbp]
  \centering
  \begin{tabular}{l || r r r r r r}
    & \multicolumn{6}{c}{$\bar{X}^{AA}_{\parallel}(\bm{q}^2) \, [\text{GeV}^2]$} \\\cline{2-7}
    ID & \multicolumn{1}{c}{(0,0,0)} & \multicolumn{1}{c}{(0,0,1)} & \multicolumn{1}{c}{(0,1,1)} & \multicolumn{1}{c}{(1,1,1)} & \multicolumn{1}{c}{(0,0,2)} & \multicolumn{1}{c}{(0,1,2)} \\
  \hline
    C-$ud3$-$s$a & $\num{0.520(29)}$ & $\num{0.533(32)}$ & $\num{0.509(83)}$ & & & \\
    C-$ud4$-$s$a & $\num{0.492(30)}$ & $\num{0.514(31)}$ & $\num{0.524(54)}$ & & & \\
    C-$ud5$-$s$a & $\num{0.465(24)}$ & $\num{0.474(29)}$ & $\num{0.376(50)}$ & & & \\
  \hline
  \hline
  C-$ud2$-$s$a-L & $\num{0.530(28)}$ & $\num{0.529(25)}$ & $\num{0.540(28)}$ & $\num{0.536(54)}$ & $\num{0.447(52)}$ & $\num{0.378(101)}$ \\
  \hline
  \hline
  M-$ud3$-$s$a & $\num{0.514(38)}$ & $\num{0.518(26)}$ & $\num{0.524(57)}$ & & & \\
  \hline
  \end{tabular}
  \caption{Same as \ref{tab:XVVParallelDataForFit} but with the data entering the fit for $\bar{X}^{AA}_{\parallel}(\bm{q}^2)$.}
  \label{tab:XAAParallelDataForFit}
\end{table}
\begin{table}[tbp]
	\centering
	\begin{tabular}{r r r r r r}
        \multicolumn{1}{c}{$c_{0} \, [\text{GeV}^2]$} & \multicolumn{1}{c}{$c_{1} \, [-]$} & \multicolumn{1}{c}{$c_{2} \, [\text{GeV}^{-2}]$} & \multicolumn{1}{c}{$\tilde{c}_{m_\pi} \, [-]$} & \multicolumn{1}{c}{$\tilde{c}_{a} \, [-]$} & \multicolumn{1}{c}{$\tilde{c}_{a\bm{q}} \, [-]$} \\
        \hline
	\multicolumn{1}{c}{$\num{0.52(7)}$} & \multicolumn{1}{c}{$\num{0.32(27)}$} & \multicolumn{1}{c}{$\num{-0.60(36)}$} & \multicolumn{1}{c}{$\num{-0.27(12)}$} & \multicolumn{1}{c}{$\num{0.65(1.78)}$} & \multicolumn{1}{c}{$\num{-1.65(2.90)}$} \\
	\hline
        \hline
        $\num{1.00}$ & $\num{-0.67}$ & $\num{-0.15}$ & $\num{-0.13}$ & $\num{-0.95}$ & $\num{0.75}$ \\
$\num{-0.67}$ & $\num{1.00}$ & $\num{-0.27}$ & $\num{0.05}$ & $\num{0.62}$ & $\num{-0.86}$ \\
$\num{-0.15}$ & $\num{-0.27}$ & $\num{1.00}$ & $\num{-0.08}$ & $\num{0.22}$ & $\num{-0.21}$ \\
$\num{-0.13}$ & $\num{0.05}$ & $\num{-0.08}$ & $\num{1.00}$ & $\num{-0.12}$ & $\num{-0.01}$ \\
$\num{-0.95}$ & $\num{0.62}$ & $\num{0.22}$ & $\num{-0.12}$ & $\num{1.00}$ & $\num{-0.75}$ \\
$\num{0.75}$ & $\num{-0.86}$ & $\num{-0.21}$ & $\num{-0.01}$ & $\num{-0.75}$ & $\num{1.00}$ \\
        \hline
	\end{tabular}
    \caption{Same as Tab. \ref{tab:XVVPerp_Multifit} but for $\bar{X}_{\parallel}^{AA}(\bm{q}^2)$. The $\chi^2/\text{d.o.f.}$ from the correlated fit is given by $0.48$.}
	\label{tab:XAAParallel_Multifit}
\end{table}

\begin{table}[tbp]
  \centering
  \begin{tabular}{l || r r r r r r}
    & \multicolumn{6}{c}{$\bar{X}^{AA}_{\perp}(\bm{q}^2) \, [\text{GeV}^2]$} \\\cline{2-7}
    ID & \multicolumn{1}{c}{(0,0,0)} & \multicolumn{1}{c}{(0,0,1)} & \multicolumn{1}{c}{(0,1,1)} & \multicolumn{1}{c}{(1,1,1)} & \multicolumn{1}{c}{(0,0,2)} & \multicolumn{1}{c}{(0,1,2)} \\
  \hline
    C-$ud3$-$s$a & $\num{1.041(58)}$ & $\num{0.419(31)}$ & $\num{0.060(27)}$ & & & \\
    C-$ud4$-$s$a & $\num{0.985(61)}$ & $\num{0.371(34)}$ & $\num{0.043(32)}$ & & & \\
    C-$ud5$-$s$a & $\num{0.930(48)}$ & $\num{0.380(31)}$ & $\num{0.033(35)}$ & & & \\
  \hline
  \hline
  C-$ud2$-$s$a-L & $\num{1.060(55)}$ & $\num{0.669(30)}$ & $\num{0.446(28)}$ & $\num{0.259(34)}$ & $\num{0.138(39)}$ & $\num{0.023(38)}$ \\  
  \hline
  \hline
  M-$ud3$-$s$a & $\num{1.028(76)}$ & $\num{0.393(36)}$ & $\num{0.074(41)}$ & & & \\
  \hline
  \end{tabular}
  \caption{Same as \ref{tab:XVVParallelDataForFit} but with the data entering the fit for $\bar{X}^{AA}_{\perp}(\bm{q}^2)$.}
  \label{tab:XAAPerpDataForFit}
\end{table}
\begin{table}[tbp]
	\centering
	\begin{tabular}{r r r r r r}
        \multicolumn{1}{c}{$c_{0} \, [\text{GeV}^2]$} & \multicolumn{1}{c}{$c_{1} \, [-]$} & \multicolumn{1}{c}{$c_{2} \, [\text{GeV}^{-2}]$} & \multicolumn{1}{c}{$\tilde{c}_{m_\pi} \, [-]$} & \multicolumn{1}{c}{$\tilde{c}_{a} \, [-]$} & \multicolumn{1}{c}{$\tilde{c}_{a\bm{q}} \, [-]$} \\
        \hline
	\multicolumn{1}{c}{$\num{1.02(14)}$} & \multicolumn{1}{c}{$\num{-3.20(91)}$} & \multicolumn{1}{c}{$\num{2.45(1.37)}$} & \multicolumn{1}{c}{$\num{-0.18(13)}$} & \multicolumn{1}{c}{$\num{0.26(1.88)}$} & \multicolumn{1}{c}{$\num{-0.96(4.75)}$} \\
	\hline
        \hline
        $\num{1.00}$ & $\num{-0.85}$ & $\num{0.72}$ & $\num{-0.05}$ & $\num{-0.95}$ & $\num{0.54}$ \\
$\num{-0.85}$ & $\num{1.00}$ & $\num{-0.98}$ & $\num{-0.05}$ & $\num{0.83}$ & $\num{-0.89}$ \\
$\num{0.72}$ & $\num{-0.98}$ & $\num{1.00}$ & $\num{0.08}$ & $\num{-0.72}$ & $\num{0.94}$ \\
$\num{-0.05}$ & $\num{-0.05}$ & $\num{0.08}$ & $\num{1.00}$ & $\num{-0.20}$ & $\num{0.16}$ \\
$\num{-0.95}$ & $\num{0.83}$ & $\num{-0.72}$ & $\num{-0.20}$ & $\num{1.00}$ & $\num{-0.59}$ \\
$\num{0.54}$ & $\num{-0.89}$ & $\num{0.94}$ & $\num{0.16}$ & $\num{-0.59}$ & $\num{1.00}$ \\
        \hline
	\end{tabular}
    \caption{Same as Tab. \ref{tab:XVVPerp_Multifit} but for $\bar{X}_{\perp}^{AA}(\bm{q}^2)$. The $\chi^2/\text{d.o.f.}$ from the correlated fit is given by $0.5$.}
	\label{tab:XAAPerp_Multifit}
\end{table}

\clearpage

\end{document}